\documentclass[journal=jacsat,manuscript=article]{achemso}
\usepackage[version=3]{mhchem}
\usepackage{xcolor}
\usepackage{graphicx}% Include figure files
\usepackage{dcolumn}% Align table columns on decimal point
\usepackage{bm}% bold math
\usepackage{hyperref}%

\author{Vikash Sharma}
\email{vikash.sharma@tifr.res.in}
\affiliation{Department of Condensed Matter Physics
and Materials Science, Tata Institute of Fundamental
Research, Mumbai 400005, India}

\author{Sitaram Ramakrishnan}
\email{niranj002@gmail.com}
\affiliation{Department of Quantum Matter, AdSE,
Hiroshima University, Higashi-Hiroshima 739-8530, Japan}
\alsoaffiliation{I-HUB Quantum Technology Foundation,
Indian Institute of Science Eduction and Research,
Pune 411008, India}

\author{S. S. Jayakrishnan}
\affiliation{Department of Mechanical Engineering,
Indian Institute of Technology Bombay, Mumbai,
MH 400076, India}

\author{Surya Rohith Kotla}
\affiliation{Laboratory of Crystallography,
University of Bayreuth, 95447 Bayreuth, Germany}

\author{Bishal Maiti}
\affiliation{Department of Condensed Matter Physics
and Materials Science, Tata Institute of Fundamental
Research, Mumbai 400005, India}

\author{Claudio Eisele}
\affiliation{Laboratory of Crystallography,
University of Bayreuth, 95447 Bayreuth, Germany}

\author{Harshit Agarwal}
\affiliation{Laboratory of Crystallography,
University of Bayreuth, 95447 Bayreuth, Germany}

\author{Leila Noohinejad}
\affiliation{P24, PETRA III,
Deutsches Elektronen-Synchrotron DESY,
Notkestrasse 85, 22607 Hamburg, Germany}

\author{M. Tolkiehn}
\affiliation{P24, PETRA III,
Deutsches Elektronen-Synchrotron DESY,
Notkestrasse 85, 22607 Hamburg, Germany}

\author{Dipanshu Bansal}
\email{dipanshu@iitb.ac.in}
\affiliation{Department of Mechanical Engineering,
Indian Institute of Technology Bombay, Mumbai,
MH 400076, India}

\author{Sander van Smaalen}
\email{smash@uni-bayreuth.de}
\affiliation{Laboratory of Crystallography,
University of Bayreuth, 95447 Bayreuth, Germany}

\author{Arumugam Thamizhavel}
\email{thamizh@tifr.res.in}
\affiliation{Department of Condensed Matter Physics
and Materials Science, Tata Institute of Fundamental
Research, Mumbai 400005, India}

\title{Room temperature charge density wave in
a tetragonal polymorph of Gd$_2$Os$_3$Si$_5$ and
study of its origin in the $RE_{2}T_{3}X_{5}$
($RE$ = Rare earth, $T$ = transition metal, $X$ = Si, Ge) series}

\abbreviations{IR,NMR,UV}


\begin{document}

\begin{tocentry}

\includegraphics{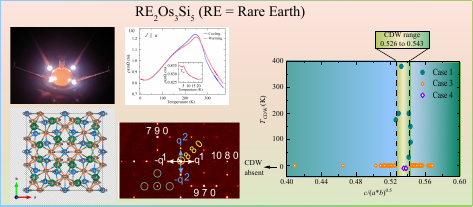}

\end{tocentry}

\begin{abstract}
Charge density wave (CDW) systems are proposed to
exhibit application potential for electronic and
optoelectronic devices.
However, CDWs often develop at cryogenic temperatures,
which hinders their applications.
Therefore, identifying new materials that exhibit
a CDW state at room temperature is crucial for the
development of CDW-based devices.
Here, we present a  non-layered tetragonal polymorph
of Gd$_2$Os$_3$Si$_5$, which exhibits a CDW state
at room temperature.
Gd$_2$Os$_3$Si$_5$ assumes the tetragonal
Sc$_2$Fe$_3$Si$_5$ structure type
with the space group $P4/mnc$.
Single-crystal X-ray diffraction (SXRD) analysis shows
that Gd$_2$Os$_3$Si$_5$ possesses an incommensurately
modulated structure with modulation wave vector
$\mathbf{q}$ = $(0.53,\, 0,\, 0)$, while the modulation
reduces the symmetry to orthorhombic $Cccm(\sigma00)0s0$.
This differs from isostructural Sm$_2$Ru$_3$Ge$_5$,
where the modulated phase has been reported to possess
monoclinic symmetry $Pm(\alpha\,0\,\gamma)0$.
Reinvestigation of Sm$_2$Ru$_3$Ge$_5$ suggests
that its modulated crystal structure can alternatively
be described by $Cccm(\sigma00)0s0$, with modulations
similar to Gd$_2$Os$_3$Si$_5$.
The temperature-dependent magnetic susceptibility
indicates an
antiferromagnetic transition at $T_{\rm N} \approx 5.5$~K.
Furthermore it shows an anomaly at around
345~K, suggesting a CDW transition at $T_{\textrm{CDW}}$
= 345~K, in agreement with
high-temperature SXRD measurements.
The temperature-dependent electrical resistivity
has a maximum at a lower temperature, that we
nevertheless identify with the CDW transition,
and that can be described
as an insulator-to-metal transition (IMT).
The calculated electronic band structure
indicates q-dependent electron-phonon couling as
dominant mechanism of CDW formation in tetragonal
Gd$_2$Os$_3$Si$_5$.
The modulated structure then indicates a major
involvement of the Si2a atom in the CDW modulations.
Compounds $RE_{2}T_{3}X_{5}$
($RE$ = rare earth, $T$ = transition metal, $X$ = Si, Ge)
have been reported with either the tetragonal
Sc$_2$Fe$_3$Si$_5$ structure type or the
orthorhombic U$_2$Co$_3$Si$_5$ structure type.
Not all of these compounds undergo
CDW phase transitions.
We find that $RE_{2}T_{3}X_{5}$  compounds will
exhibit a CDW transition, if the condition
$0.526~<c/\sqrt{ab}~<~0.543$ is satisfied.
\end{abstract}

\maketitle

\clearpage

\section{\label{sec:gd2os3si5_introduction}Introduction}

A charge-density-wave (CDW) is a modulation of the
electronic density of a metallic, crystalline material,
which is accompanied by a periodic lattice distortion
(PLD) of the atomic positions.
\cite{wilsonja1974a,wilsonja1975a,grunerg1988a,teng2022discovery,liu2016charge,meier2021catastrophic,bugaris2017charge}.
The wavelength of the CDW/PLD represents an independent
period in the system, such that, in general, the material
becomes an aperiodic crystal.
A CDW can develop upon cooling a sample from the
normal state,\cite{grunerg1988a,bugaris2017charge}
or by applying pressure on it.\cite{zhao2018pressure}
In the CDW phase, the electronic structure of a
metallic system becomes modified.
A gap opens up in the electronic density of states,
which is often reflected in the physical properties,
such as the electrical resistivity and magnetic susceptibility,
and in spectroscopic
experiments~\cite{grunerg1988a,chen2017charge,bugaris2017charge,teng2022discovery,Wandel_science2022}.
There are various techniques to probe the CDW state,
such as X-ray diffraction
(XRD)\cite{PhysRevX.11.031050,Wandel_science2022},
scanning tunnelling microscopy (STM) \cite{Gao2019real,teng2022discovery},
transmission electron microscopy (TEM) \cite{Meyer2011},
angle resolved X-ray photoelectron spectroscopy (ARPES) \cite{teng2022discovery},
nuclear magnetic resonance (NMR)\cite{wu2011magnetic},
Raman spectroscopy~\cite{loret2019} and
ultrafast pump-probe experiments~\cite{chenry2017a}.
A metal-to-insulator transition (MIT), obvious
in the electrical resistivity, together with the
appearance of superlattice reflections in XRD are
the typical characteristic features pointing
toward a CDW state \cite{bugaris2017charge, meier2021catastrophic,
ramakrishnan2023a, ramakrishnan2020a, ramakrishnan2021a}.

Compounds $RE_{2}T_{3}X_{5}$
($RE$ = Rare earth, $T$ = transition metal, $X$ = Si, Ge)
are known to exhibit polymorphism,
crystallizing in either the Sc$_2$Fe$_3$Si$_5$ structure
type with the tetragonal space group $P4/mnc$ or the
U$_2$Co$_3$Si$_5$ structure type with the orthorhombic
space group $Ibam$ \cite{brown2023a}.   CDW transitions have been observed for compounds with either structure type.
For example, Ramakrishnan \textit{et al.}
\cite{ramakrishnan2020a,ramakrishnan2021a,ramakrishnan2023a}
reported CDW transitions in three orthorhombic compounds $RE_2$Ir$_3$Si$_5$,
with transition temperatures $T_{\rm CDW}$ = 90~K for $RE$ = Ho,
$T_{\rm CDW}$ = 150~K for $RE$ = Er,
and $T_{\rm CDW}$ = 200~K for $RE$ = Lu.
Moreover, Ho$_2$Ir$_3$Si$_5$ shows coupling between CDW ordering and magnetism~\cite{ramakrishnan2023a}.
XRD studies pointed towards zigzag chains of Ir atoms being responsible
for the stabilization of the CDW in these compounds
\cite{ramakrishnan2023a,ramakrishnan2020a, ramakrishnan2021a}.
Bugaris \textit{et al.}~\cite{bugaris2017charge}
have reported CDW transitions in the tetragonal compounds $RE_2$Ru$_3$Ge$_5$ ($RE$= Pr, Sm, Dy), with $T_{\rm CDW}$ = 200~K for $RE$ = Pr, and $T_{\rm CDW}$ = 175~K for $RE$ = Sm.
Stabilization of the CDWs was attributed to a PLD on the zigzag chains of Ge~\cite{bugaris2017charge}.
Interestingly, one such compound Sm$_{2}$Ru$_{3}$Ge$_{5}$
shows CDW transitions
in both the orthorhombic and tetragonal polymorphs
\cite{bugaris2017charge, kuo2020a, sokkalingamr2023a}.

There are other compounds in the series $RE_{2}T_{3}X_{5}$  that do not exhibit CDW transitions, despite being isostructural to the orthorhombic or tetragonal structure types.  For example, orthorhombic compounds $RE_2$Ir$_3$Si$_5$ containing lighter rare-earth elements $RE$ = La, Ce or Gd, do not show CDW transitions~\cite{Yogesh2004}.
Tetragonal compounds $RE_2$Ru$_3$Ge$_5$ do not exhibit CDW order for  rare-earth elements $RE$ other than those mentioned above \cite{bugaris2017charge}.
Furthermore, tetragonal compounds $RE_2$Re$_3$Si$_5$ ($RE$ = Ce, Pr, Ho) do not exhibit CDW transitions in the temperature range $2-300$~K~\cite{sanki2022a,sharma2022a}.

In one of the earlier studies,
Gd$_2$Os$_3$Si$_5$ has been reported with
the orthorhombic U$_2$Co$_3$Si$_5$ structure type,
albeit with disorderly occupation of one of the two
Os sites by Os/Si, resulting in the composition
Gd$_2$Os$_{2.82(2)}$Si$_{5.14(2)}$  \cite{mykhalichko2015a}.
However, no physical properties have been reported so far.
Here, we report the synthesis and characterization
of a new polymorph of Gd$_2$Os$_3$Si$_5$.
It crystallizes in the tetragonal Sc$_2$Fe$_3$Si$_5$ structure type
with space group $P4/mnc$.
We have discovered an incommensurate CDW state in this compound,
which is characterized by the modulation wave vector
$\mathbf{q}$ = $(0.53,\, 0,\, 0)$.
The onset of the CDW transition appears to be somewhere between 400 to 345~K,
and thus shows the potential for applications in electronic devices~
\cite{parson2003a,fu2020room,chen2017charge,fragkos2019room,feng2020achieving,liu2016charge}.
Analysis by SXRD reveals that the CDW modulation
is of orthorhombic symmetry with superspace group $Cccm(\sigma00)0s0$,
while the average crystal structure remains tetragonal.
Using electronic structure calculations, we evaluated several mechanisms
of CDW formation, including Fermi surface nesting and hidden nesting,
and find that the wave vector-dependent electron-phonon
coupling possibly controls the CDW formation.
The electrical resistivity $\rho(T)$ depicts an insulator
to metal transition (IMT) at around 250~K along with a
hysteresis during cooling and warming cycles in insulating as well as metallic regimes.
This insulating behaviour is in line with observed
weak superlattice reflections in SXRD at 300~K, pointing
a CDW state even at room temperature.
The magnetic susceptibility $\chi(T)$ reveals an
antiferromagnetic ordering at $T\rm_N$ $\sim$ 5.5~K,
and in the paramagnetic state it exhibits
hysteresis where the CDW transition is observed.
The previous results of $RE_{2}T_{3}X_{5}$ compounds
have revealed that the dimensionality plays a
key role in the formation of the CDW phase~
\cite{ramakrishnan2023a,ramakrishnan2020a, ramakrishnan2021a,
Yogesh2004,bugaris2017charge,sanki2022a,sharma2022a}.
A criterion is proposed, which is based on the lattice
parameters according to the value of $c/\sqrt{ab}$,
where $a$, $b$ and $c$ are the lattice parameters
of the tetragonal or orthorhombic structure.
We found that $0.526~<c/\sqrt{ab}~<$~0.543 needs to be
fulfilled for the existence of a CDW state in a compound
of the series of $RE_{2}T_{3}X_{5}$.

\section{\label{sec:gd2os3si5_experimental_methods}%
Experimental methods and computational details}

\subsection{\label{sec:gd2os3si5_crystal_growth}%
Crystal growth and physical properties}

\begin{figure}[ht]
\includegraphics[width=80mm]{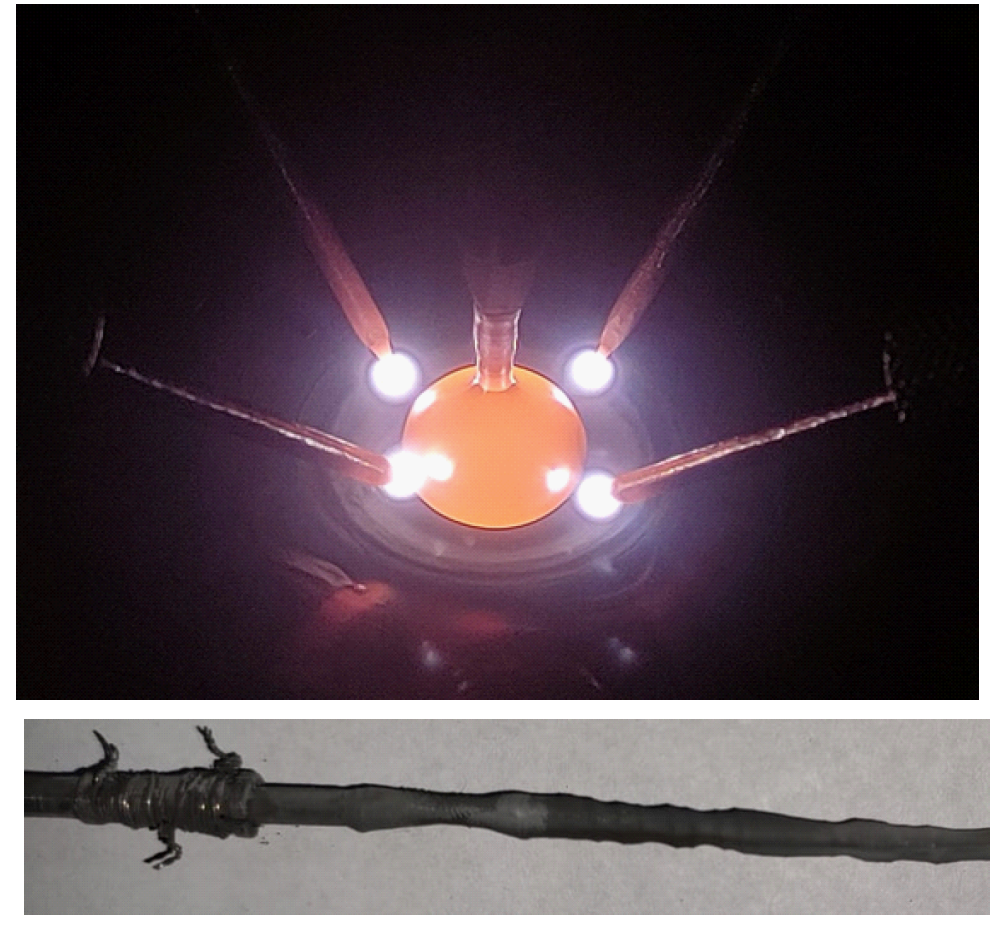}
\caption{\label{fig:gd2os3si5_crystal_fig1}%
Photograph of the crystal growth process in a tetra-arc
furnace and the final pulled ingot.}
\end{figure}

We have prepared a Gd$_2$Os$_3$Si$_5$ single crystal using the Czochralski method
in a tetra-arc furnace (Technosearch Corporation, Japan) under ultra-pure argon atmosphere.
The high purity starting elements of Gd, Os, and Si were taken in the molar ratio $2:\,3:\,5$ and melted to prepare a polycrystalline ingot of composition Gd$_2$Os$_3$Si$_5$.
The melting process was repeated multiple times to make the sample homogeneous.
A seed crystal was cut from the polycrystalline ingot and used in
the Czochralski growth of Gd$_2$Os$_3$Si$_5$.  The seed crystal was gently inserted
into the molten solution of Gd$_2$Os$_3$Si$_5$ and pulled rapidly at a rate of 50 mm/h.
A necking process was made so that a single crystalline grain is selected for
the growth of a single crystal.  Once the steady state condition is achieved the crystal was pulled at a rate of 10 mm/h.
A snapshot of the crystal growth process is shown in Fig.~\ref{fig:gd2os3si5_crystal_fig1}.
We obtained a long, rod-shaped ingot with an average
diameter of 4~mm and a length of 50~mm as shown in the bottom part of Fig.~\ref{fig:gd2os3si5_crystal_fig1}.
A Laue diffraction pattern in back-reflection geometry
confirmed the single crystalline nature of the pulled
ingot as well as its tetragonal symmetry.
The structure did not change upon annealing at 1273 K
for four days.
The composition was confirmed through the energy
dispersive X-ray analysis (EDX) attachment in the
field emission scanning electron microscope (FESEM)
and the anisotropic studies were carried out using
commercial equipment like physical property measurement system (PPMS)
and superconducting quantum interference device-vibrating sample magnetometer (SQUID-VSM).
Detailed explanation for the experimental methods
regarding EDX and physical properties are given
in the supporting information \cite{gd2os3si5suppmat2023a}.

\subsection{\label{sec:gd2os3si5_t_dependent_sxrd}%
Temperature-dependent X-ray diffraction}

%X-ray powder diffraction at room temperature.
Initially, powder X-ray diffraction (PXRD) was
measured at room temperature on an in-house
Panalytical X-ray diffractometer, employing
Cu--K$_{\alpha}$ radiation.
Rietveld refinements of the PXRD data
were performed with
the  FullProf software \cite{rodriguez_carvajal1993a}.
Satellite reflections could not be observed
in PXRD nor could any lattice distortion away
from tetragonal be observed.
This result already suggests that the
incommensurate CDW modulation is a small
deviation from $P4/mnc$ symmetry.

%Single-crystal diffraction
Single-crystal X-ray diffraction (SXRD)
experiments were performed on
a small piece of crystal of dimensions
of $0.15\times 0.07\times 0.1$ mm$^3$,
that was acquired by crushing the large,
annealed single crystal.
Complete data sets of SXRD data were measured at temperatures
of 300, 200, 100 and 20~K with a Pilatus 1M CdTe detector.
After this cooling run, an additional data set was measured at 230~K.
High-temperature SXRD was measured on a second
single crystal, with dimensions:
$0.098~\times~0.026~\times~0.108$~mm$^3$.
The experiment was done on the same
diffractometer but with a different detector (Lambda CdTe 7.5~M).
Data sets were collected at 400 and 450~K. Both SXRD measurements
were carried out employing synchrotron radiation of $\lambda_{P24} = 0.5$~\AA{}.
Details of both PXRD and SXRD measurements are given
in sections S1 and S2 in
the supporting information \cite{gd2os3si5suppmat2023a}

The EVAL15 software suite \cite{schreursamm2010a} was used
for processing the SXRD data.
SADABS \cite{sheldrick2008} was used for scaling and
absorption correction.
The latter employed Laue symmetry 4/$mmm$ for
the periodic phase and $mmm$ for the CDW phase.
As the crystal structure in the CDW phase is
incommensurately modulated we had to use
the superspace approach \cite{van2007incommensurate,wagner2009a,stokesht2011a}
to index and integrate the data.
The reflection file produced was imported into the
software JANA2006 for structure refinement \cite{petricekv2016a, petricekv2014a}.
Table \ref{tab:gd2os3si5_cdw_crystalinfo} gives
the crystallographic information at
450 and 400~K (periodic phase),
and at 300 and 200~K (incommensurate phase).
An unusual criterion of observability was chosen,
in order to account for the incorrect estimate of
the standard uncertainties in the integrated
intensities by the software EVAL15.
For details regarding
the processing of SXRD data
refer to section S3 in the supporting information~\cite{gd2os3si5suppmat2023a}.
%
%*********************************Table1*************************************************
\begin{table}[th]
\caption{\label{tab:gd2os3si5_cdw_crystalinfo}%
Crystallographic data of
Gd$_2$Os$_3$Si$_5$ at 450, 400  300  and 200~K.
At temperatures 450 and 400~K, the crystal is in the periodic phase
and at lower temperatures 300 and 200~K the crystal is in the incommensurate CDW phase.}
	\scriptsize
	\centering
	\begin{tabular}{ccccc}
		\hline
		Temperature (K) &450 & 400  & 300 & 200  \\
		Crystal system & Tetragonal & Tetragonal & Orthorhombic& Orthorhombic  \\
		Space/Superspace group & $P4/mnc$ & $P4/mnc$ & $Cccm(\sigma00)0s0$  &
		$Cccm(\sigma00)0s0$ \\
		Space/Superspace group No. \cite{stokesht2011a} & 128& 128 & {66.1.15.8} & {66.1.15.8}   \\
		$a$ (\AA{}) &10.7163(2) & 10.7122(2) &15.1389(3) &15.1279(3)  \\
		$b$ (\AA{}) &10.7163 & 10.7122 &15.1380(2)    &15.1271(2)     \\
		$c$ (\AA{}) &5.7030(1) & 5.7019(3) &5.6998(2)    & 5.6993(2)  \\
		Volume (\AA{}$^3$) & 654.93(2) & 654.30(4) & 1306.25(5) & 1304.24(5)  \\
		Wave vector, $\mathbf{q}_x$ & -& -&0.5381(2) & 0.5374(3)  \\
		$Z$ &4 &4 & 8 & 8 \\
		Wavelength (\AA{}) & 0.50000 &0.50000 & 0.50000 &0.50000  \\
		Detector distance (mm) & 100& 100 &110 &110  \\
		$2\theta$-offset (deg) &0 &0 &0 &0 \\
		$\chi$-offset (deg) &-60 & -60 &-60 & -60  \\
		Rotation per image (deg) & 1 & 1  & 1 & 1 \\
		$(\sin(\theta)/\lambda)_{max}$ (\AA{}$^{-1}$) & 0.712376 & 0.717144 &0.683977& 0.683955 \\
		Absorption, $\mu$ (mm$^{-1}$) & 31.462 &31.492 & 31.549 & 31.597  \\
		T$_{\rm min}$, T$_{\rm max}$ & 0.1479, 0.2602 & 0.1426, 0.2540 & 0.0216, 0.0534 & 0.0212, 0.0524  \\
		Criterion of observability & $I>0.75\sigma(I)$ & $I>0.75\sigma(I)$ & $I>0.75\sigma(I)$ & $I>0.75\sigma(I)$ \\
		Number of $(m = 0)$ reflections \\
		measured & 9250& 9448 &  5329  & 5235  \\
		unique (obs/all) & 542/545 & 556/558 &912/962 &908/968  \\
		Number of $(m = 1)$ reflections \\
		measured & -& -& 50157 & 50078  \\
		unique (obs/all) & -& -& 1529/7648 & 2104/7630  \\
		$R_{int}$ $(m = 0)$ (obs/all) & 5.27/5.27& 5.15/5.15 &0.0468/0.0469 &0.0460/0.0461 \\
		$R_{int}$ $(m = 1)$ (obs/all) & -&- &0.0661/0.2700 &0.0765/0.2178 \\
		No. of parameters & 31 & 31 &98 &98 \\
		$R_{F }$ $(m = 0)$  (obs) & 0.0178 & 0.0185 &0.0204 &0.0237 \\
		$R_{F }$ $(m = 1)$ (obs) & -&- &0.0603 &0.0493 \\
		$wR_{F }$ $(m = 0)$ (all) &0.0216 & 0.0223&0.0242 &0.0272 \\
		$wR_{F }$ $(m = 1)$ (all) & -&- &0.0897 &0.0610 \\
		$wR_{F }$ all (all) & 0.0216 & 0.0223 &0.0256 &0.0284 \\
		GoF (obs/all) & 1.56/1.56 & 1.62/1.62 &0.83/0.45 &0.81/0.49 \\
		$\Delta\rho_{min}$, $\Delta\rho_{max}$(e \AA$^{-3}$) & -3.77, 4.11 & -3.71, 3.68 &
		-2.53, 2.15 &-2.62, 1.71  \\ \\
		\hline
	\end{tabular}
\end{table}
%*********************************End Table1********************************************
%
Crystallographic tables of other temperatures,
and details regarding SXRD data collection and
data processing are given in the
Supporting Information (SI)~\cite{gd2os3si5suppmat2023a}.

\subsection{\label{sec:gd2os3si5_electronic_structure_calc}%
	Electronic structure calculations}

\begin{figure}[ht]
\centering
\includegraphics[trim=0.0cm 0.0cm 0.0cm 0cm, clip=true,width=0.90\textwidth]{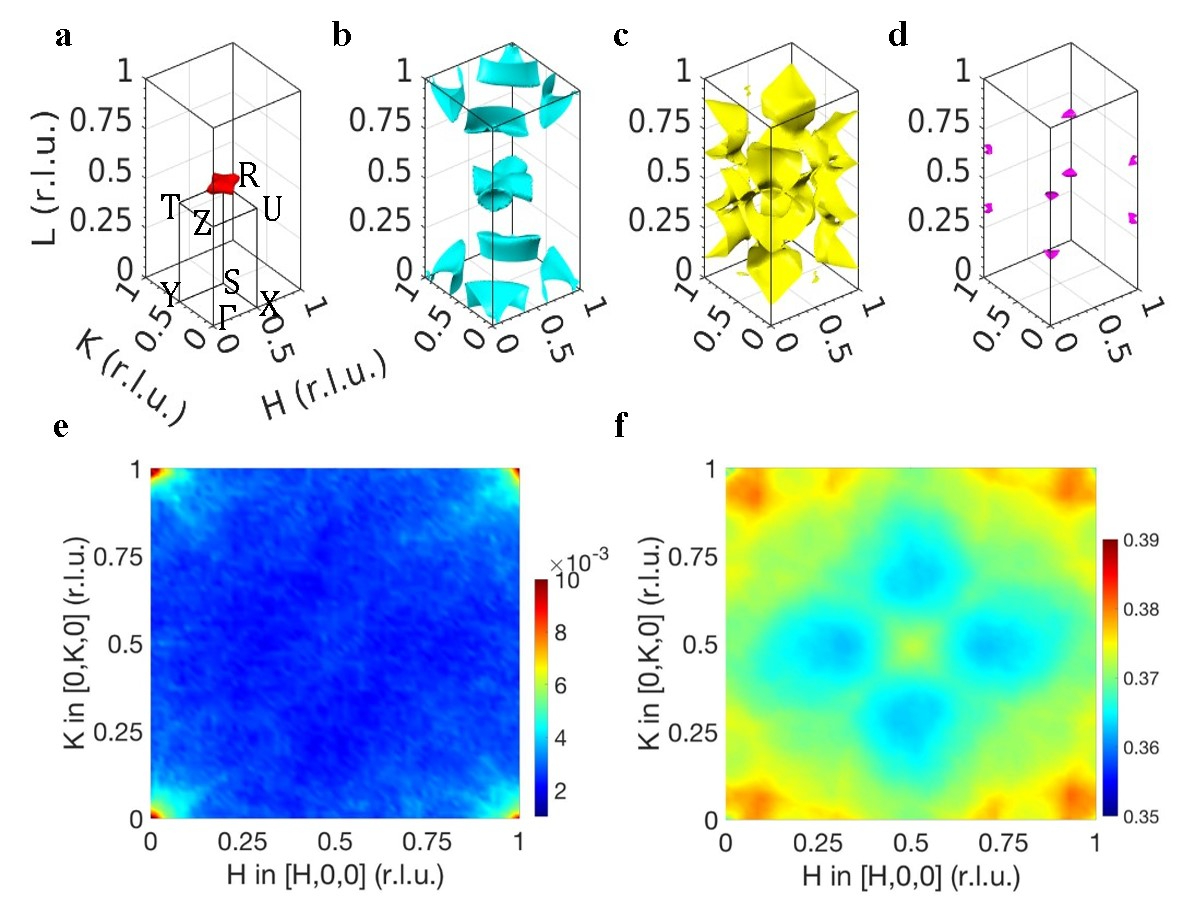}
\caption[Fermi surface topology for Gd$_2$Os$_3$Si$_5$]
{\label{fig:gd2os3si5_fermisurfaces}%
(a-d) Fermi surface topology for Gd$_2$Os$_3$Si$_5$. Fermi surface is reconstructed from the four bands that cross $E_{\rm F}$. Separate plots of the Fermi surface from different bands are for easy visualization. High-symmetry $k$-points of the Brillouin zone are shown in panel (a). r.l.u. refers to the reciprocal lattice units. (e,f) The imaginary and real part of Lindhard susceptibility, i.e., Im$\{\chi_0({\bf q},\omega)\}$ and Re$\{\chi_0({\bf q},\omega)\}$ for $\omega = 0$, in the [$H$,$K$,0] scattering plane. Colors represent the magnitude of Re$\{\chi_0({\bf q},\omega)\}$ and Im$\{\chi_0({\bf q},\omega)\}$ in arbitrary units. No divergence is visible at experimental CDW wave vectors, thus ruling out any role of Fermi surface nesting and hidden nesting in inducing the CDW.}
\end{figure}

Electronic structure calculations were performed
within the framework of density functional theory (DFT) as implemented in the Vienna ab initio Simulation Package~\cite{Kresse1996}.
We have computed the electronic structure for the
tetragonal structure of Gd$_2$Os$_3$Si$_5$
(space group $P4/mnc$).
This periodic crystal structure contains
$40$ atoms (8 Gd atoms, 12 Os atoms and 20 Si atoms)
in the primitive unit cell with lattice parameters
$a$~=~10.706, $b$~=~10.706 and $c$~=~5.701~\AA{}
(compare to the $C$-centered setting in
Table \ref{tab:gd2os3si5_cdw_crystalinfo}).
We used a $3\times 3\times 6$ $\Gamma$-centered
Monkhorst-Pack electronic $k$-point mesh,
with the plane-wave cut-off energy of 500~eV
for the converged results (change in the total electron
free energy is $10^{-3}$~eV/atom when
increasing the mesh size to $4~\times~4\times~8$
or higher grid density).
The convergence criteria for the self-consistent
electronic loop was set to $10^{-8}$~eV.
The projector-augmented-wave potentials explicitly
included $5d^1 4f^7$ states for Gd,
$5d^76s^1$ states for Os and
$3s^23p^2$ for Si as valence electrons.
Gd$_2$Os$_3$Si$_5$ undergoes an antiferromagnetic (AFM)
transition at $T_{\rm N}\sim 5.5$~K,
%($\mu_{eff}=7.91\mu_{\rm B}$/Gd)
as determined from the electrical resistivity and  magnetic susceptibility measurements.
Hence, we performed spin-polarized calculations.
Spin-orbit coupling (SOC) has minimal influence on ground-state properties; consequently, we did not include SOC in our calculations except for initial testing.
We used the generalized gradient approximation (GGA) in the revised Perdew-Burke-Ernzerhof (PBE) parametrization~\cite{refPBE} with or without a Hubbard correction. To account for the localized $f$ orbitals of Gd atoms and $d$ orbitals for Os atoms, GGA+$U$ calculations were used~\cite{Dudarev_1998} with onsite Coulomb interaction $U$ = 6.0~eV for $f$ orbitals of Gd atoms and $U$ = 2.0~eV for $d$ orbitals of Os atoms~\cite{hou2015lattice}.
On increasing the $U$ value of
Os from 0 to 2 eV, the band structure shifts
by nearly 0.1 eV (see Fig. S3 in the SI).
The shift has a little influence on the Fermi surface
topology and electron susceptibility, as described below.
During the relaxation, atomic positions were
optimized until forces on all atoms were smaller
than 1~meV~\AA{}$^{-1}$.  The Fermi surface topology has been calculated by first computing the electron bands on a $3~\times~3~\times~4$ grid during the self consistent field (SCF) step. Subsequently, we employed the software WANNIER90 (v3.1.0)\cite{pizzi2020wannier90} to compute maximally localized Wannier functions on a 100-orbital basis.
The basis included 40~$d$ orbitals of eight Gd atoms and 60 $d$ orbitals of twelve Os atoms in the unit cell, as both significantly contributed to the electronic bands near the Fermi level ($E_{F}$).
To confirm the accuracy of Wannierization, we compared the electronic band structure obtained from the Wannier functions with the direct DFT calculations (see Fig.S4 in SI).
We constructed the Fermi surface by evaluating electronic bands on a $70\times70\times70$ grid in the entire Brillouin zone using WANNIER90 and plotted it using in-house scripts as shown in Fig.~\ref{fig:gd2os3si5_fermisurfaces}.
%
%**********************************************Fig2**********************
%****************************************Fig2********************************
%

The electronic susceptibility for a non-interacting Fermi gas,
that excludes electron-electron correlations but includes Pauli's exclusion principle can be calculated using the Lindhard function.
The first-order perturbative linear response of this gas to a external perturbation of frequency $\omega$ and wave vector $\mathbf{q}$ is given by~\cite{dressel_gruner_2002}:
\begin{equation}\label{eqn:gd2os3si5_susceptibility1}
	\chi_0({\bf q},\omega) = \lim_{\gamma\rightarrow 0}\frac{e^2}{\Omega}
	\sum\limits_{{\bf k}}\sum\limits_{l,l'}
	\frac{f\left(\epsilon_{{\bf k+q},l'}\right) -
		f\left(\epsilon_{{\bf k},l}\right)}
	{\epsilon_{{\bf k+q},l'} - \epsilon_{{\bf k},l} - \hbar\omega - i\gamma}
	|\langle {\bf k+q},l'|V|{\bf k},l\rangle|^2,
\end{equation}
where $e$ is the electron charge,
$\Omega$ is the system volume,
$f\left(\epsilon_{k,l}\right)$ denotes the electron's Fermi distribution function at energy $\epsilon$ for a given wave vector $\mathbf{k}$ and a band index $l$,
$|\langle \mathbf{k+q},l'|V|\mathbf{k},l\rangle|^2$ is the matrix transition element between states $\langle \mathbf{k+q},l'|$ and $|\mathbf{k},l\rangle$, and $\gamma$ is a infinitesimal constant.
The first and second summations represent the sum over all $k$ states and all bands $l$ in first Brillouin zone, respectively.
The real and imaginary part of Lindhard susceptibility are subsequently calculated from Eq.~\eqref{eqn:gd2os3si5_susceptibility1} as follows:
\begin{equation}\label{eqn:gd2os3si5_real_susceptibility}
	\begin{split}
		{\rm Re}\{\chi_0({\bf q},\omega)\} = \lim_{\gamma\rightarrow 0}
		\frac{e^2}{\Omega}\sum\limits_{{\bf k}}\sum\limits_{l,l'}
		\frac{\left(\epsilon_{{\bf k+q},l'} - \epsilon_{{\bf k},l} - \hbar\omega\right)\left(f\left(\epsilon_{{\bf k+q},l'}\right)
			- f\left(\epsilon_{{\bf k},l}\right)\right)}
		{{\left(\epsilon_{{\bf k+q},l'} - \epsilon_{{\bf k},l} -
				\hbar\omega\right)^2 + \gamma^2}}\times \\
		|\langle {\bf k+q},l'|V|{\bf k},l\rangle|^2
	\end{split}
\end{equation}
\begin{equation}\label{eqn:gd2os3si5_imaginary_susceptibility}
	{\rm Im}\{\chi_0({\bf q},\omega)\} = \lim_{\gamma\rightarrow 0}\frac{e^2}{\Omega}\sum\limits_{{\bf k}}\sum\limits_{l,l'}\frac{-\gamma\left(f\left(\epsilon_{{\bf k+q},l'}\right) - f\left(\epsilon_{{\bf k},l}\right)\right)}{{\left(\epsilon_{{\bf k+q},l'} - \epsilon_{{\bf k},l} - \hbar\omega\right)^2 + \gamma^2}}|\langle {\bf k+q},l'|V|{\bf k},l\rangle|^2.
\end{equation}
If $\epsilon_{\mathbf{k}} = \epsilon_{\mathbf{k+q}} = E_{\rm F}$, Eq.~\eqref{eqn:gd2os3si5_imaginary_susceptibility} reduces to the Fermi surface nesting (FSN) function, where $E_{\rm F}$ is the Fermi energy.
To evaluate the expressions in Eqs.~\eqref{eqn:gd2os3si5_real_susceptibility} and~\eqref{eqn:gd2os3si5_imaginary_susceptibility}, certain practical aspects need to be considered.
We include bands that lie close to $E_{\rm F}$ and contribute to the Lindhard susceptibility, i.e., four bands that cross $E_{\rm F}$.
We assume transition matrix elements to be unity, in-line with earlier studies~\cite{johannes2008fermi,zhu2015classification,roy2021quasi,pathak2022orbital}. Because of a finite size $k$-grid, we use a finite but small value of $|\gamma| = 1$~meV to broaden the functions with width comparable to the energy difference between the discrete $k$ points (although because of this we sacrifice some fine structure).
Re$\{\chi_0(\mathbf{q},\omega)\}$ and
Im$\{\chi_0(\mathbf{q},\omega)\}$ for $\omega = 0$ in
the ($H$,$K$,0) scattering plane is shown in Fig.~\ref{fig:gd2os3si5_fermisurfaces}(e,f).

\section{\label{sec:gd2os3si5_results_discussion}%
Results and Discussion}

\subsection{\label{sec:gd2os3si5_incommensurate_structure}%
Incommensurately modulated structure of the CDW state}

\begin{figure}[ht]%
\includegraphics[width=0.8\textwidth]{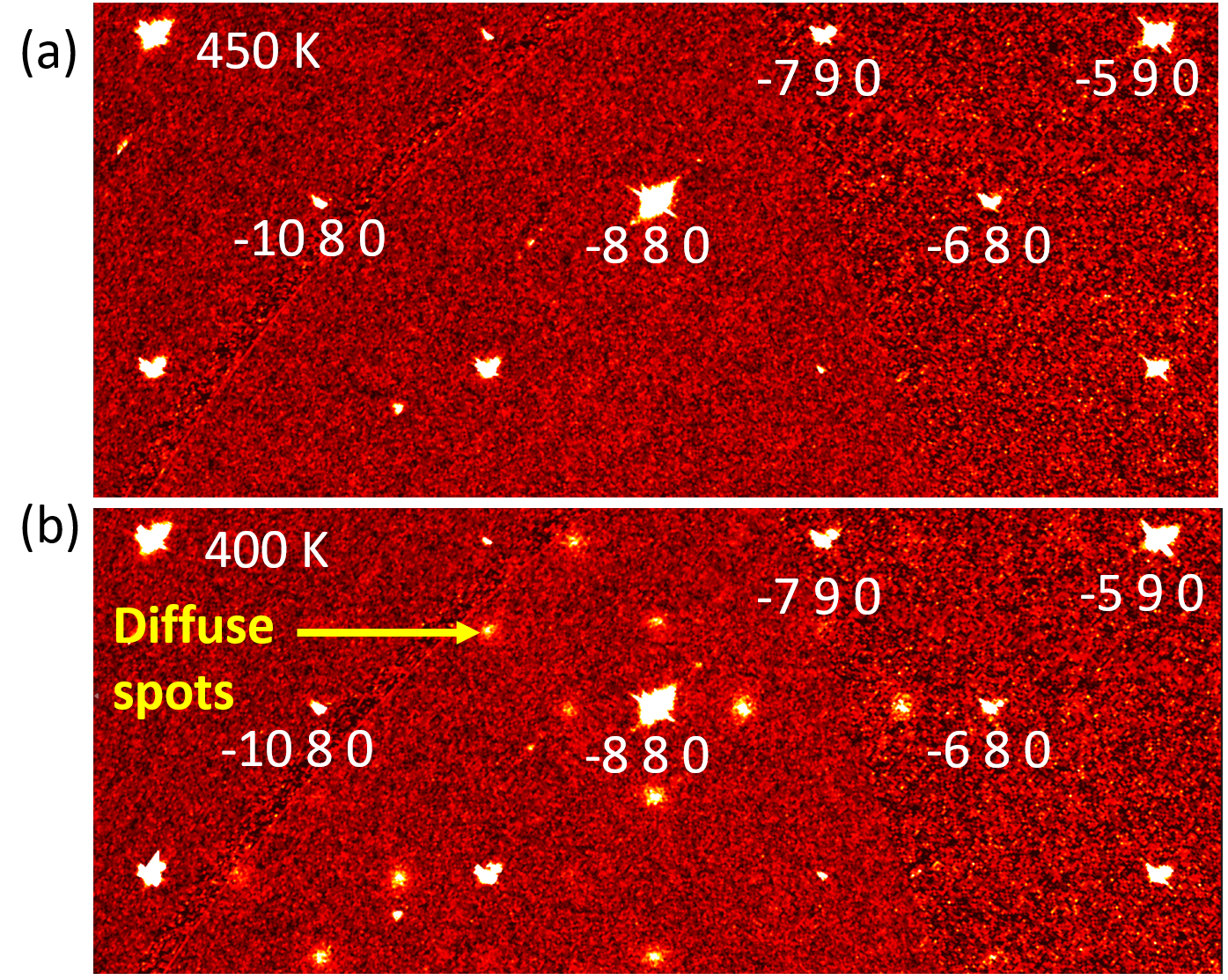}
\caption{\label{fig:Gd2Os3Si5_HT_unwarp}%
Excerpts of the reconstructed reciprocal layers
$(h\,k\,0)$ of diffraction, for temperatures of
(a) 450~K and (b) 400~K.
Indices are given for several main reflections
according to the $C$-centered setting.
These diffuse spots indicated by a yellow arrow, become satellite reflections
at lower temperatures, as shown in Fig. \ref{fig:gd2os3si5_unwarp}.
}
\end{figure}

\begin{figure}[ht]
\includegraphics[width=80mm]{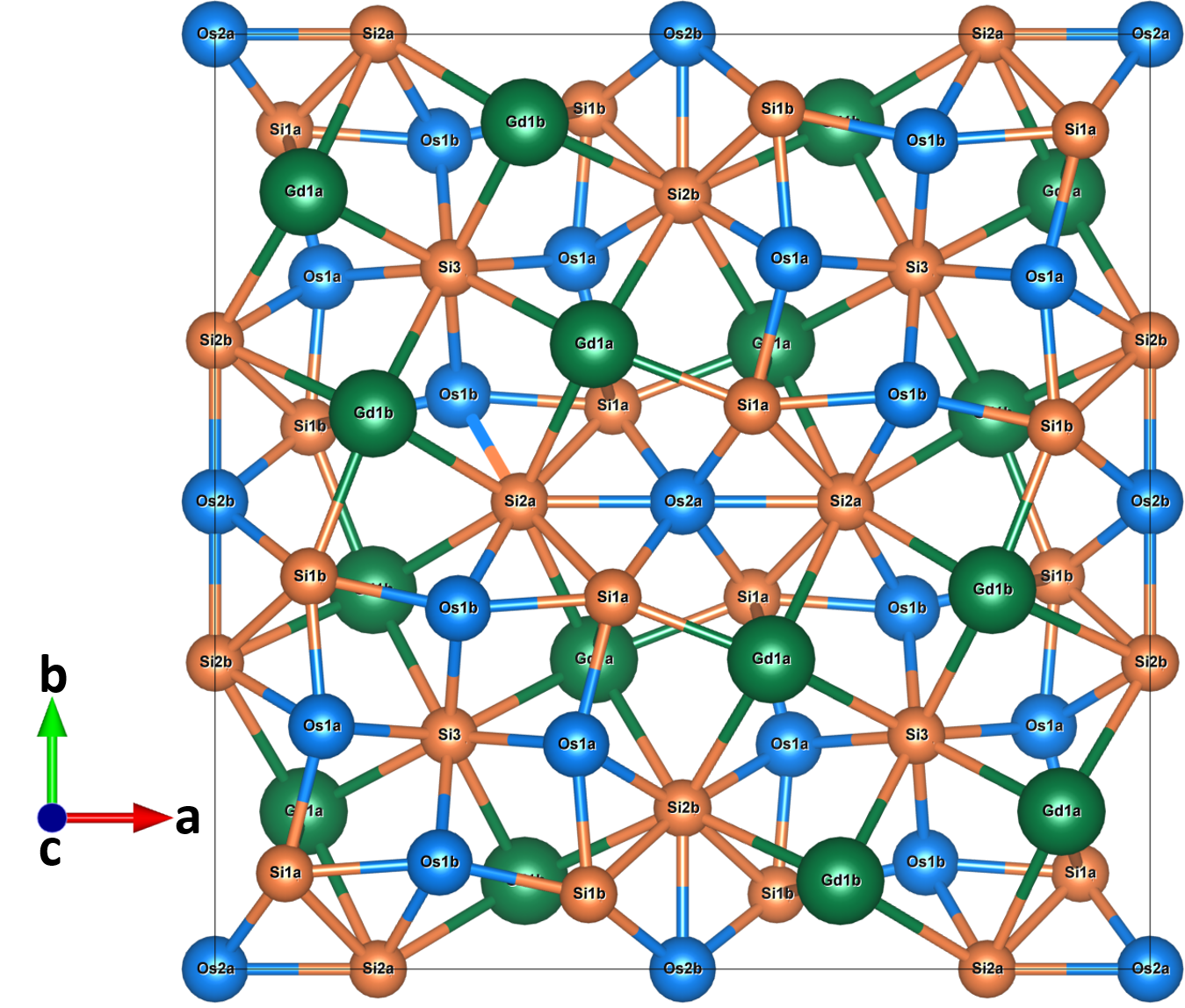}
\caption{\label{fig:gd2os3si5_unit_cell}%
Projection onto the $(ab)$-plane
of the average crystal structure of
Gd$_{2}$Os$_{3}$Si$_{5}$ at 300~K, depicting
one $C$-centered unit cell.
Large green spheres correspond to gadolinium,
blue spheres of intermediate sizes correspond
to osmium and small orange spheres depict silicon.
Atom numbers are indicated.}
\end{figure}

According to the present SXRD experiments,
Gd$_2$Os$_3$Si$_5$ adopts the tetragonal
Sc$_2$Fe$_3$Si$_5$ structure type for temperatures
450 and 400~K (Figs. \ref{fig:Gd2Os3Si5_HT_unwarp}
and\ \ref{fig:gd2os3si5_unit_cell}).
%
%******************************Figure 3******************
%****************************End Figure 3***************************
%
%
%***********************Figure4*************************************
%***********************Figure4*************************************
%
%(Section \ref{sec:gd2os3si5_t_dependent_sxrd}).
The observed scattering near the satellite positions
in the SXRD at 400~K  is diffuse scattering
(Fig. \ref{fig:Gd2Os3Si5_HT_unwarp}) and it indicates CDW fluctuation above $T_{\rm CDW}$.
Diffuse scattering at temperatures above $T_{\rm CDW}$ has been
observed for other CDW systems too \cite{pougetjp2024a}.
The symmetry of the periodic phase is given by
the space group $P4/mnc$,
much like $RE_2$Re$_3$Si$_5$ ($RE$= Ce, Pr, Ho) \cite{sanki2022a, sharma2022a}
and $RE_2$Ru$_3$Ge$_5$ ($RE$= Dy, Sm, Pr) \cite{bugaris2017charge}.
Structure refinements led to an excellent fit
to the SXRD data (Table \ref{tab:gd2os3si5_cdw_crystalinfo}).
Structural parameters are provided in Table S5
in the Supplementary Material.\cite{gd2os3si5suppmat2023a}

\begin{figure}[tb]
\includegraphics[width=0.8\textwidth]{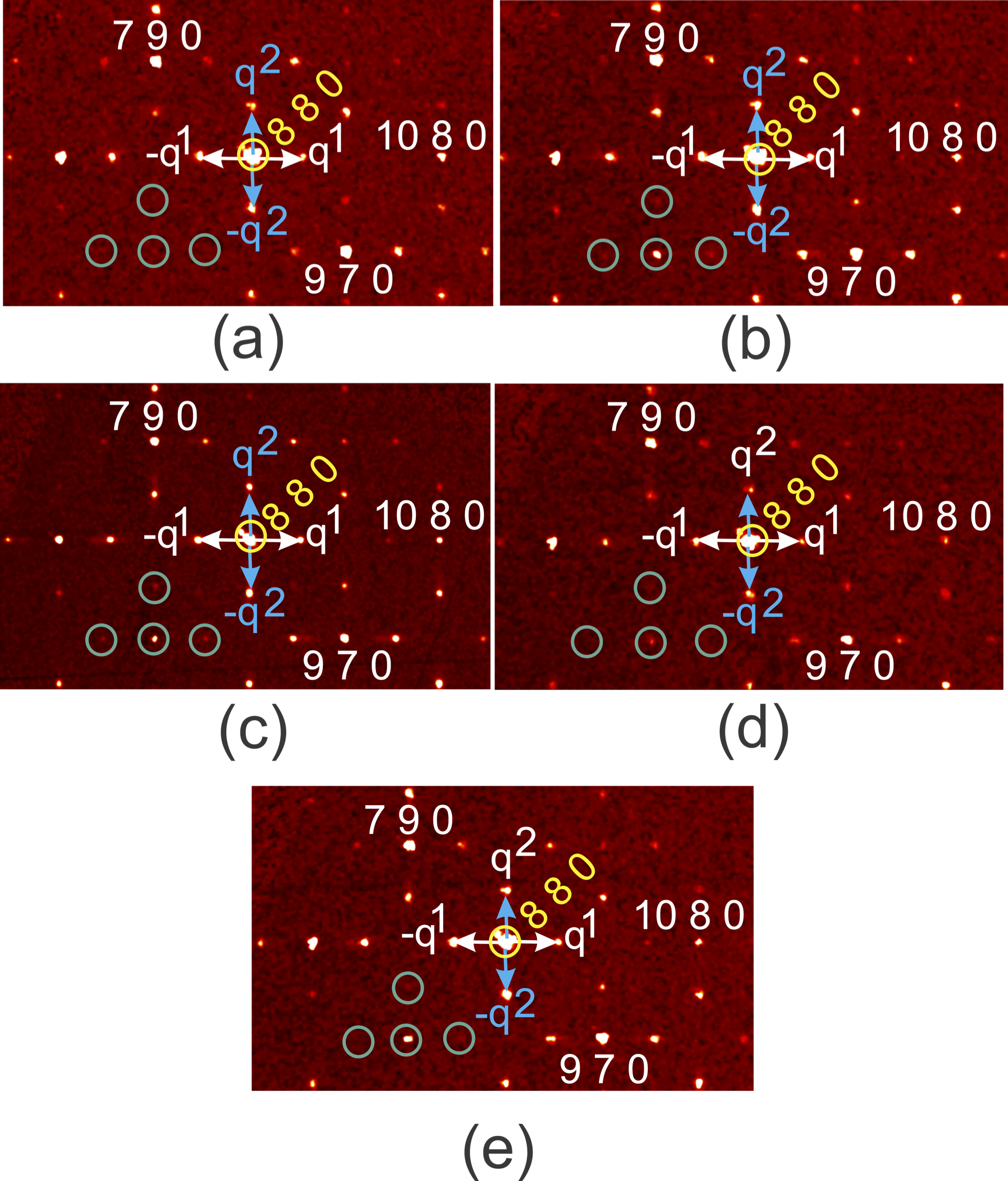}
\caption{\label{fig:gd2os3si5_unwarp}%
Reconstructed reciprocal layers
$(h\,k\,0)$ of SXRD measured at temperatures of
(a) 300~K,
(b) 230~K,
(c) 200~K,
(d) 100~K, and
(e) 20~K.
Indices are given for several main reflections and
satellite reflections according to the $C$-centered setting.
The yellow circles highlight the main reflection $(8\,8\,0)$.
It is surrounded by four superlattice reflections,
whose positions are defined by
$\pm\mathbf{q}^1$ and $\pm\mathbf{q}^2$
(Eq. \protect~\ref{eqn:gd2os3si5_q1_q2_vectors}).
$\mathbf{q}^1$ and $\mathbf{q}^2$ are related by twinning.
Green circles indicate weakening of reflections at around 100~K
(see Fig. \ref{fig:gd2os3si5_lattpar_vs_temp}(d) for details.)
}
\end{figure}

For temperatures 300~K and below,
SXRD has indicated the presence
of Bragg reflections in addition to those of the
Sc$_2$Fe$_3$Si$_5$ structure type.
These reflections can be indexed as
incommensurate satellite reflections
(Fig.~\ref{fig:gd2os3si5_unwarp}).
%
%********************Figure5*****************************
%*******************END Figure5**************************
%
They are evidence
for an incommensurately modulated structure.
This structure has been determined by
structure refinements within the superspace approach
(Table \ref{tab:gd2os3si5_cdw_crystalinfo}).

We did not observe any distortion of the
tetragonal lattice of the average structure.
Furthermore, the observed modulation wave vectors
obey the tetragonal symmetry according to
\begin{align}
\label{eqn:gd2os3si5_q1_q2_vectors}
\mathbf{q}^{1} &= (\sigma_1,\,\sigma_1,\,0) \nonumber \\
\mathbf{q}^{2} &= (-\sigma_1,\,\sigma_1,\,0)
\end{align}
with $\sigma_1 = 0.2691(3)$ at $T = 300$ K.
Here, the fourfold rotation transforms
$\mathbf{q}^{1}$ into $\pm\mathbf{q}^{2}$, and
$\mathbf{q}^{2}$ into $\mp\mathbf{q}^{1}$.
An immediate choice for the symmetry of the
incommensurately modulated structure would thus
be a (3+2)-dimensional [(3+2)D] superspace group
based on the tetragonal space group $P4/mnc$
of the average structure together with the two modulation
wave vectors of Eq. \ref{eqn:gd2os3si5_q1_q2_vectors}.
However, these models failed to describe the
intensities of the satellite reflections.
See model A in
Table \ref{tab:gd2os3si5_refmod_compare}.
%
%*********************************Table2*******************
\begin{table}[ht]
\tiny
\scriptsize
\caption{\label{tab:gd2os3si5_refmod_compare}%
Crystallographic data of three structure models
for the incommensurate CDW phase at $T=200$ K.
Criterion of observability of Bragg reflections
is: $I>0.75\sigma(I)$.}
\centering
\begin{tabular}{ccccc}
\hline
Model  & A  &  B & C &  D \\
Crystal system & Tetragonal & Tetragonal & Orthorhombic & Orthorhombic \\
Superspace group &
\parbox[c]{32mm}{$P4/mnc(\sigma\sigma0)0000$\\ \hfill$(-\sigma\sigma0)0000$} &
\parbox[c]{32mm}{$P4/m(\sigma_{1}\sigma_{2}0)0000\\(-\sigma_{2}\sigma_{1}0)0000$} &
\parbox[c]{32mm}{$Pnnm(\sigma_{1}\sigma_{2}0)000\\(-\sigma_{1}\sigma_{2}0)000$} &
$Cccm(\sigma00)0s0$ \\
\parbox[c]{26mm}{Superspace\\ group No.\cite{stokesht2011a}} & {128.2.68.5} & {83.2.57.1} & {58.2.50.13} & {66.1.15.8} \\
$a$ (\AA{}) & 10.7042(2)   & 10.7042(2) & 10.7042(2)   & 15.1279(3) \\
$b$ (\AA{}) &10.7042    & 10.7042  & 10.7042(2) &   15.1271(2)      \\
$c$ (\AA{}) &5.6993(2)    & 5.6993(2)  & 5.6995(3)  & 5.6993(2)   \\
Volume (\AA{}$^3$) & 653.03(5) & 653.03(5)& 653.05(5) & 1304.24(5) \\
Wave vector, $\mathbf{q}^1$ & $(0.2691(3), 0.2691(3), 0)$
& $(0.2691(3), 0.2691(3), 0)$ & $(0.2691(3), 0.2691(3), 0)$  & $(0.5374(3), 0, 0)$ \\
Wave vector, $\mathbf{q}^2$ & $(-0.2691(3), 0.2691(3), 0)$
& $(-0.2691(3), 0.2691(3), 0)$ & $(-0.2691(3), 0.2691(3), 0)$  & - \\
$Z$ & 4 & 4 & 4 & 8 \\
Laue symmetry & $4/mmm$ & $4/m$ & $mmm$ & $mmm$ \\
$R_{int}$(all)(obs)   & 0.0500 & 0.0461 & 0.0446 & 0.0467 \\
$R_{int}(m = 0)$(obs) & 0.0486 & 0.0453 & 0.0437 & 0.0460 \\
$R_{int}(m = 1)$(obs) & 0.0879 & 0.0737 & 0.0758 & 0.0765 \\
No. of parameters & 66 & 131 & 130 & 98 \\
$R_{F }$ $(m = 0)$  (obs) &0.0233 &0.0248 &0.0227 &0.0237 \\
$R_{F }$ $(m = 1)$ (obs) &0.5315 &0.0591 &0.0583 &0.0493 \\
$wR_{F }$ $(m = 0)$ (all) &0.0256 &0.0276 &0.0263 &0.0272 \\
$wR_{F }$ $(m = 1)$ (all) &0.6061 &0.0680 &0.0655 &0.0610 \\
$wR_{F }$ all (all) &0.1036 &0.0291 &0.0279 &0.0284\\
GoF (obs/all) &3.18/2.12 &0.84/0.49 &0.84/0.49 &0.81/0.49\\
$\Delta\rho_{min}$, $\Delta\rho_{max}$(e \AA{}$^{-3}$) &
-19.43, 19.41 & -2.77, 2.41 & -3.05, 1.74 &-2.62, 1.71 \\ \\
\hline
\end{tabular}
\end{table}
%*******************End Table2*****************************
%

Model B refers to a lowering of the average symmetry
towards tetragonal $P4/m$, which is a subgroup of
index two of $P4/mnc$.
This symmetry lowering introduces the possibility
of twinning with two domains \cite{parson2003a}.
The modulation remains 2D, but the modulation
wave vectors now have two independent unrestricted
components, although experimental values indicate
them to be equal.
Crystal structure refinements led to a good
fit to the diffraction data.
See model B in Table \ref{tab:gd2os3si5_refmod_compare}.

Another subgroup of $P4/mnc$ is the orthorhombic
space group $Pnnm$.
The modulation remains 2D.
Model C with the $(3+2)$D superspace group symmetry
based on $Pnnm$ provides a fit of similar quality as
model B (Table \ref{tab:gd2os3si5_refmod_compare}).

As final option we have considered structures with
1D modulations, where each domain is modulated with
either $\mathbf{q}^{1}$ or $\mathbf{q}^{2}$.
This choice of symmetry is motivated by the
failure to observe mixed-order satellite reflections
at $\mathbf{q}^{1}\pm\mathbf{q}^{2}$.
The highest such possible symmetry is the
orthorhombic subgroup $Cccm$ of $P4/mnc$.
Structure refinements with a (3+1)D superspace group
based on the average-structure symmetry $Cccm$ led
to the best fit to the diffraction data.
See model D in
Table \ref{tab:gd2os3si5_refmod_compare}.
An additional argument in favor of model D is
that the lower $R$ values are obtained for a model
with a considerably smaller number of parameters
for model D than for models B or C
(Table \ref{tab:gd2os3si5_refmod_compare}).

Table S4 in in the Supplementary Material~\cite{gd2os3si5suppmat2023a}
gives an overview of $R$ values for a total
of 26 possible superspace symmetries.
It reveals that the best fit to the SXRD data
is indeed found for model D with $Cccm(\sigma00)0s0$.
Non-centrosymmetric subgroups lead to nearly
equal $R$ values, while the number of parameters
is 1.6--2 times larger.
The present SXRD data thus indicate a centrosymmetric
CDW in Gd$_2$Os$_3$Si$_5$.

Based on these considerations, we propose for the
incommensurately modulated phase of Gd$_2$Os$_3$Si$_5$
a structure model involving a 1D modulation and
Superspace symmetry $Cccm(\sigma\,0\,0)0s0$.
Structural parameters are provided in Tables S6 and S7
in the Supplementary Material \cite{gd2os3si5suppmat2023a}.

\begin{figure}[t]
\includegraphics[width=80mm,height=80mm,keepaspectratio]{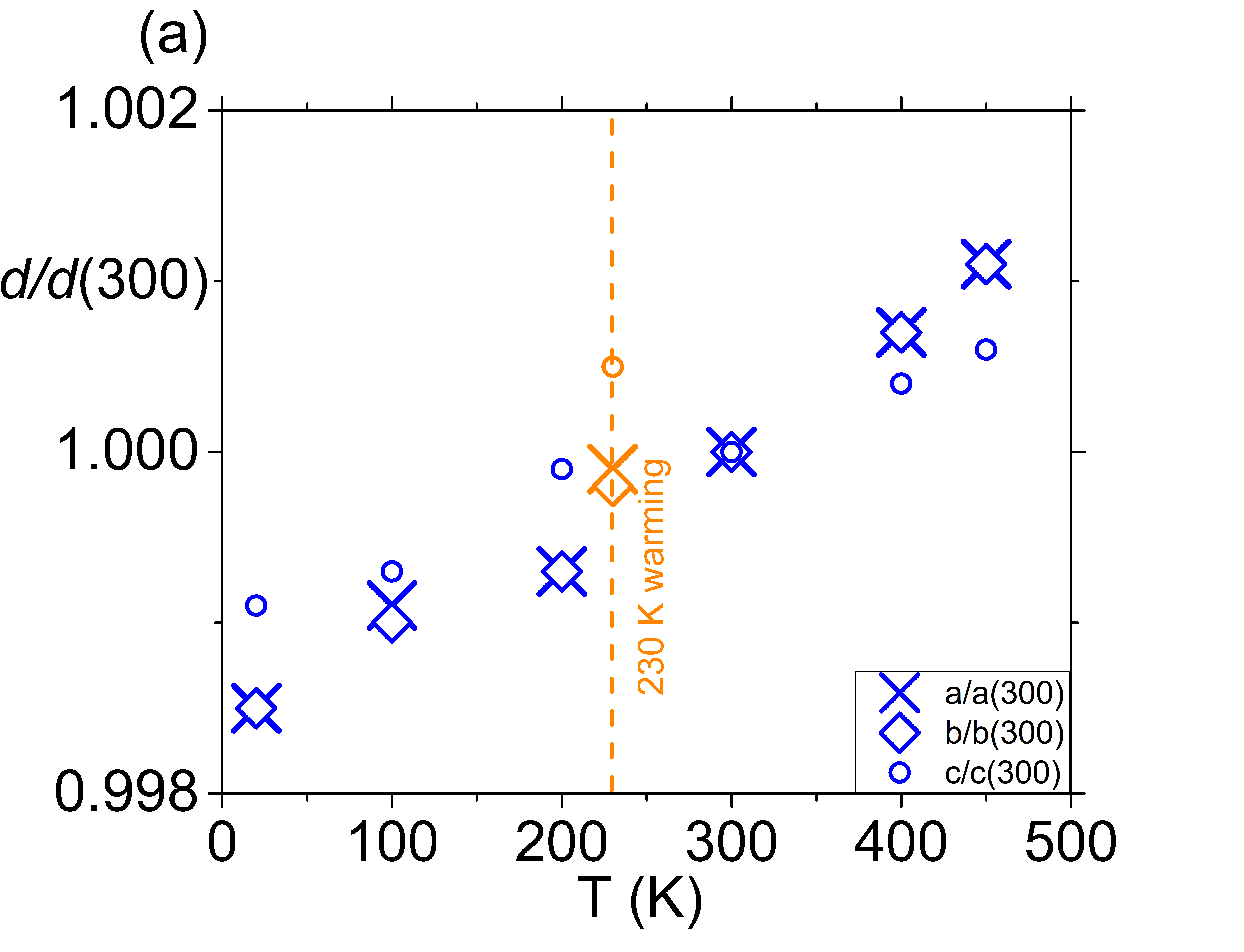}
\includegraphics[width=80mm,height=80mm,keepaspectratio]{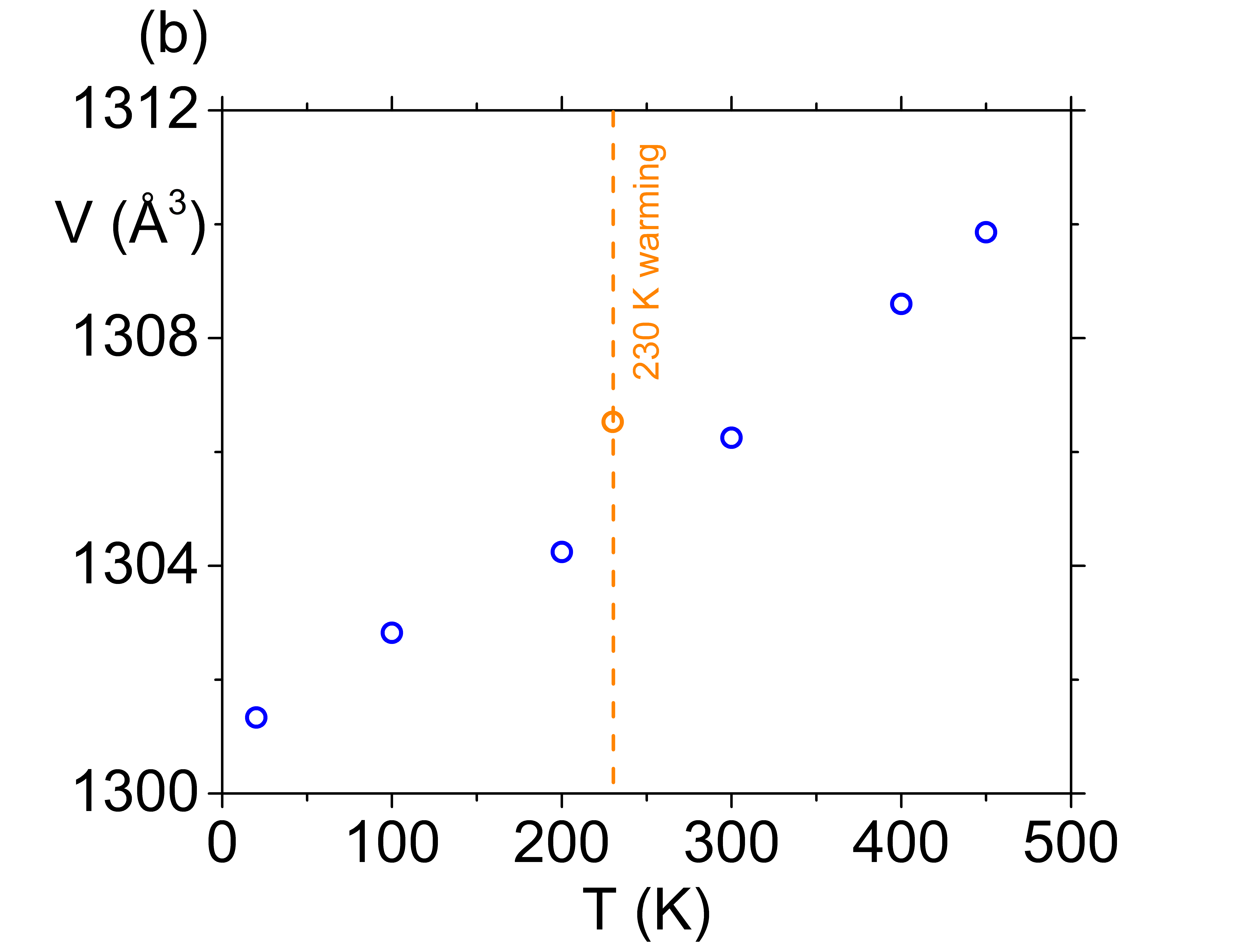}
\\
\includegraphics[width=80mm,height=80mm,keepaspectratio]{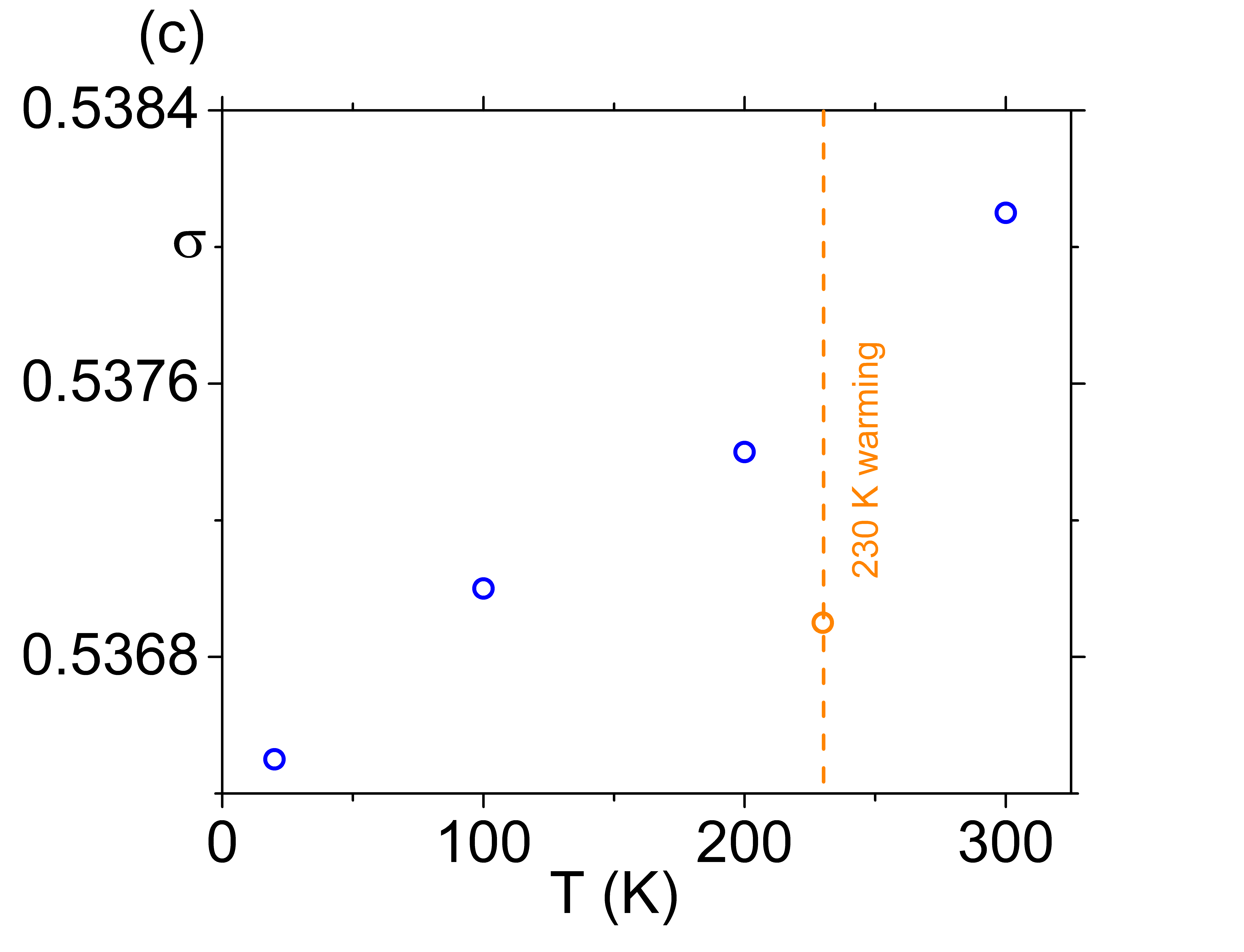}
\includegraphics[width=80mm,height=80mm,keepaspectratio]{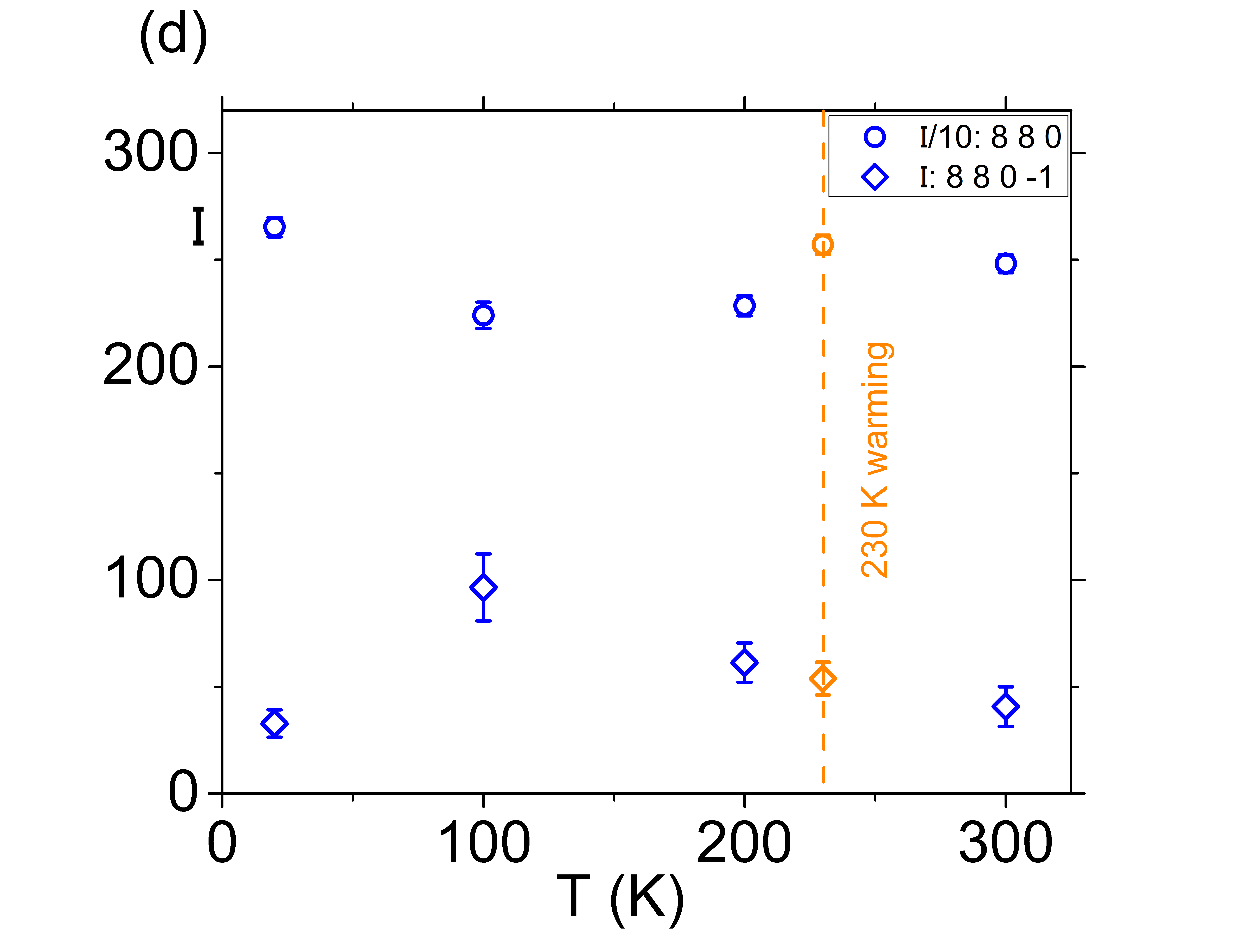}
%\includegraphics[width=80mm,height=80mm,keepaspectratio]{lattice.eps}
%\includegraphics[width=80mm,height=80mm,keepaspectratio]{volume.eps}
%\\
%\includegraphics[width=80mm,height=80mm,keepaspectratio]{sigma.eps}
%\includegraphics[width=80mm,height=80mm,keepaspectratio]{intensity.eps}
%\includegraphics[width=0.9\textwidth]{Fig6.png}
\caption{\label{fig:gd2os3si5_lattpar_vs_temp}%
(a) values of the lattice parameters relative to
their values at 300~K; with a(300)~=~15.1389(3),
b(300)~=~15.1380(2), and c(300)~=~5.6998(2)~(\AA{}).
(b) Unit cell volume.
(c) Component $\sigma$ of $\mathbf{q}$.
(d) Intensity of main reflection $(8\,8\,0)$ and
satellite reflection $(8\,8\,0\,{-}1)$ as a
function of temperature.
The data were recorded during cooling,
except those measured at 230~K (marked by a
dashed line) that were recorded during heating
of the sample.
450 and 400~K are in the tetragonal phase,
while lower temperatures from 300~K to 20~K are in the
orthorhombic CDW phase.}
\end{figure}

At all measured temperatures within the
range 20–300 K,
the lattice parameters of the average structure obey tetragonal symmetry restrictions (Fig. \ref{fig:gd2os3si5_lattpar_vs_temp}),
although the modulated structure clearly is orthorhombic.
%
%*********************************Figure6***********************************************
%*******************Figure6***************************
%
Only a few cases are known, for which the lowering of
the crystal symmetry is entirely due to the
modulation of atomic positions, including the CDW compounds
CuV$_2$S$_4$, EuAl$_4$, Sm$_2$Ru$_3$Ge$_5$ and SrAl$_4$
\cite{ramakrishnan2019a,ramakrishnan2022a,bugaris2017charge,ramakrishnan2024a}.

We have not observed a lock-in transition down to 20~K
[Fig. \ref{fig:gd2os3si5_lattpar_vs_temp}(c)].
In most cases, the intensities of superlattice
reflections should increase with decreasing temperature,
because of an increasing strength of the modulation.
However, we presently find a reduction of the
intensity of satellite reflections below 100~K
[Fig. \ref{fig:gd2os3si5_lattpar_vs_temp}(d)].
This observation represents a similar trend as
reported for Sm$_2$Ru$_3$Ge$_5$ \cite{bugaris2017charge}.
%%%%%%%%%%%%%%%%%

Isostructural Gd$_2$Os$_3$Si$_5$ and Sm$_2$Ru$_3$Ge$_5$ compounds
feature  similar modulation wave vectors.
Bugaris \textit{et al.}\cite{bugaris2017charge}
have reported a modulated crystal structure for
Sm$_2$Ru$_3$Ge$_5$, which has monoclinic symmetry
with the (3+1)D superspace group $Pm(\alpha\,0\,\gamma)0$
(Table \ref{tab:gd2os3si5_sm2ru3ge5}).
%
%******************Table3 ****************************
\begin{table}[th]
\scriptsize
\small
\caption{\label{tab:gd2os3si5_sm2ru3ge5}%
Crystallographic information on the
monoclinic and orthorhombic
structure models for the incommensurate
CDW phase of Sm$_2$Ru$_3$Ge$_5$ at 100~K.
%Structure refinements have been performed
%against the SXRD published by
%Bugaris \textit{et al.}\protect~\cite{bugaris2017charge}.
}
\centering
\begin{tabular}{ccc}
\hline
Chemical formula & \multicolumn{2}{c}{Sm$_2$Ru$_3$Ge$_5$} \\
Superspace group & $^{a}Pm(\alpha\,0\,\gamma)0$ &$^{b}Cccm(\sigma\,0\,0)0s0$\\
No.\cite{stokesht2011a} & 6.1.2.1 & 66.1.15.18 \\
$a$ (\AA{}) & 10.9955(2) & 15.5491(2)  \\
$b$ (\AA{}) &5.7823(3)   &  15.5493(3) \\
$c$ (\AA{})  &10.9942(3) &   5.7823(3) \\
$\alpha$ (deg)  &  90 &  90   \\
$\beta$ (deg) &  90  &  90   \\
$\gamma$ (deg)  &    90 & 90  \\
Volume (\AA{}$^3$)  & 699.00(4) & 1398.31(4) \\
\multicolumn{3}{l}{Modulation wave vector} \\
$\sigma_1$ & $0.219(5)$ & $0.4380(5)$ \\
$\sigma_2$ & $0$ & $0$ \\
$\sigma_3$ & $0.219(4)$ & $0$ \\
$Z$ & 4 & 8 \\
Criterion of observability & $I>2\sigma(I)$ & $I>2\sigma(I)$\\
Independent &  17468 & 17112 \\
No. of parameters &617  &97 \\
$R_{F }$ $(m = 0)$  (obs/all) &0.0607/0.0609  &0.0574/0.0576\\
$R_{F }$ $(m = 1)$ (obs/all) &0.1207/0.2730 & 0.1528/0.3058 \\
$wR_{F }$ $(m = 0)$ (all) &0.1704 &0.1097  \\
$wR_{F }$ $(m = 1)$ (all) &0.3011 &0.2211 \\
$wR_{F }$ all (all) &0.1752 &0.1211 \\
\hline
\end{tabular}\\
$^a$Information taken from
Bugaris \textit{et al.}\protect~\cite{bugaris2017charge}.\\
$^b$Structure refinements against SXRD data kindly supplied by
Bugaris \textit{et al.}\protect~\cite{bugaris2017charge}.
\end{table}
%*****************End Table3*************************
%
Here, the average-structure space group $Pm$
is a subgroup of $P4/mnc$, where the mirror
perpendicular to the tetragonal axis is preserved
in the monoclinic symmetry.
Motivated by their similar average crystal
structures, we have refined a structure model with
orthorhombic $Cccm(\sigma\,0\,0)0s0$ symmetry
against the diffraction data from
Bugaris \textit{et al.}\cite{bugaris2017charge}
for Sm$_2$Ru$_3$Ge$_5$
(Table \ref{tab:gd2os3si5_sm2ru3ge5}).
Whereas the fit to the main reflections is
the same for the monoclinic and orthorhombic
structure models, the fit to the satellite
reflections is worse for orthorhombic symmetry.
However, the monoclinic model has 617
parameters, whereas the orthorhombic model
only needs 97 parameters.
Secondly, neither $R_F^{obs}(m=1) = 0.121$
nor $R_F^{obs}(m=1) = 0.15$ represent good
fits to the satellite reflection intensities.
Considering the massive increase in number
of parameters upon lowering the symmetry
from $Cccm$ towards $Pm$, we believe that
there is no conclusive evidence for either
structure model.
A structure with superspace symmetry
$Cccm(\sigma\,0\,0)0s0$ thus is quite well
possible for the CDW phase of Sm$_2$Ru$_3$Ge$_5$.

\subsection{\label{sec:gd2os3si5_location_of_cdw}%
Location of the CDW in Gd$_2$Os$_3$Si$_5$}

The tetragonal crystal structure with symmetry
$P4/mnc$ involves six crystallographically
independent atoms,
Gd1, Os1, Os2, Si1, Si2 and Si3 respectively.
The basic structure of the incommensurately
modulated CDW structure possesses the symmetry $Cccm$.
Employing group--subgroup relations, one
finds that five out of these six sites split
into two independent positions, enumerated by
Gd1a, Gd1b, Os1a, Os1b, Os2a, Os2b, Si1a, Si1b,
Si2a, Si2b and Si3.
Their positions in the $C$-centered unit cell
are shown in Fig. \ref{fig:gd2os3si5_unit_cell}.
Deviations from tetragonal symmetry are very
small for the refined basic-structure coordinates
(Table S6 in \cite{gd2os3si5suppmat2023a}).

\begin{figure}[t]
\includegraphics[width=80mm]{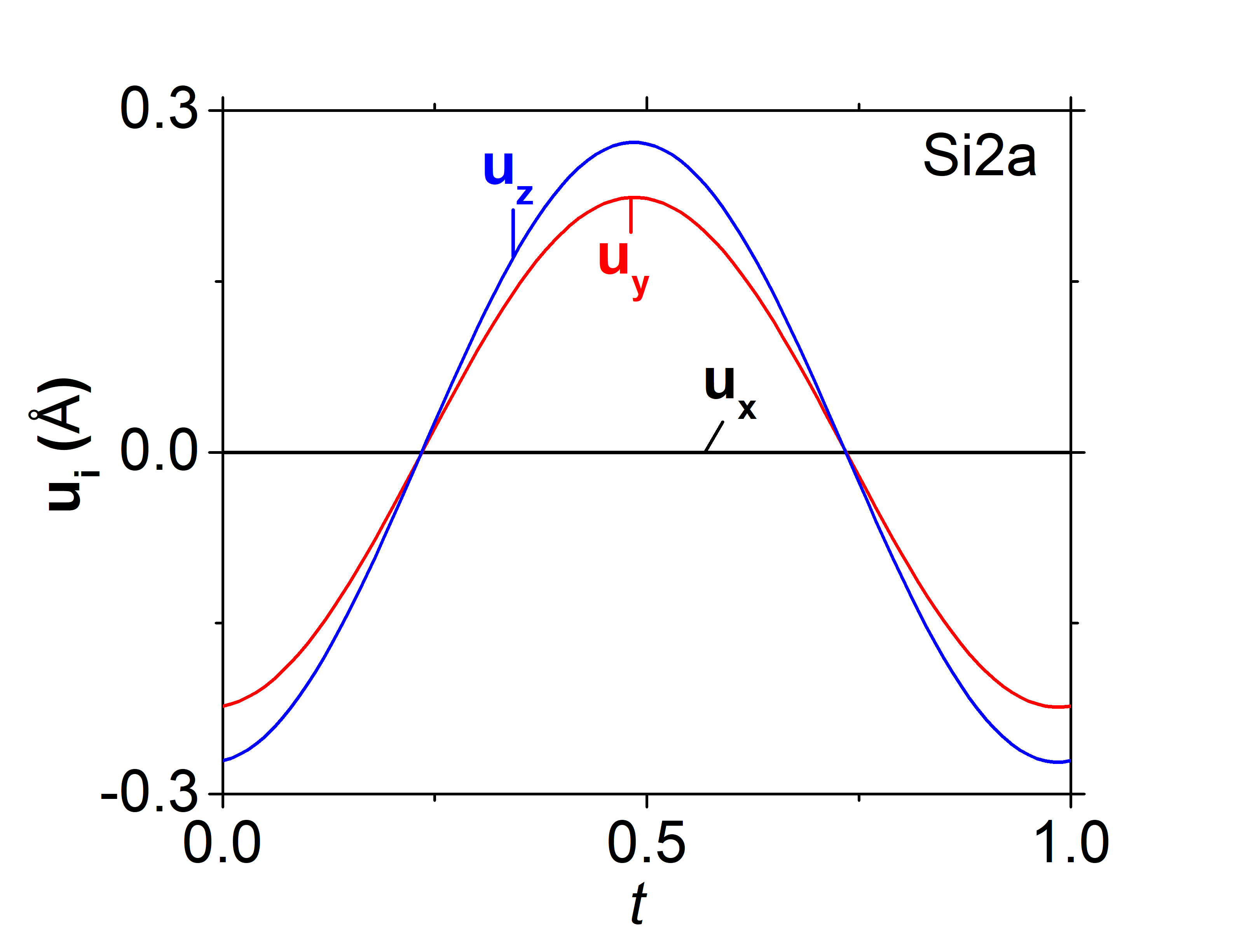}
\caption{\label{fig:gd2os3si5_disp_si2a}%
$t$-plot of displacement modulation of
Si2a at 200~K.}
\end{figure}

Inspection of the list of modulation amplitudes
reveals that by far the largest displacements are
those of the Si2a atom
(Table S7 in \cite{gd2os3si5suppmat2023a}).
Specifically, large displacements of Si2a are
found along $\mathbf{b}$ and $\mathbf{c}$
(Fig. \ref{fig:gd2os3si5_disp_si2a}),
%
%**********************Figure7***********************************
%************************************************************
%******************************Fig8************************
\begin{figure}[t]
\includegraphics[width=80mm]{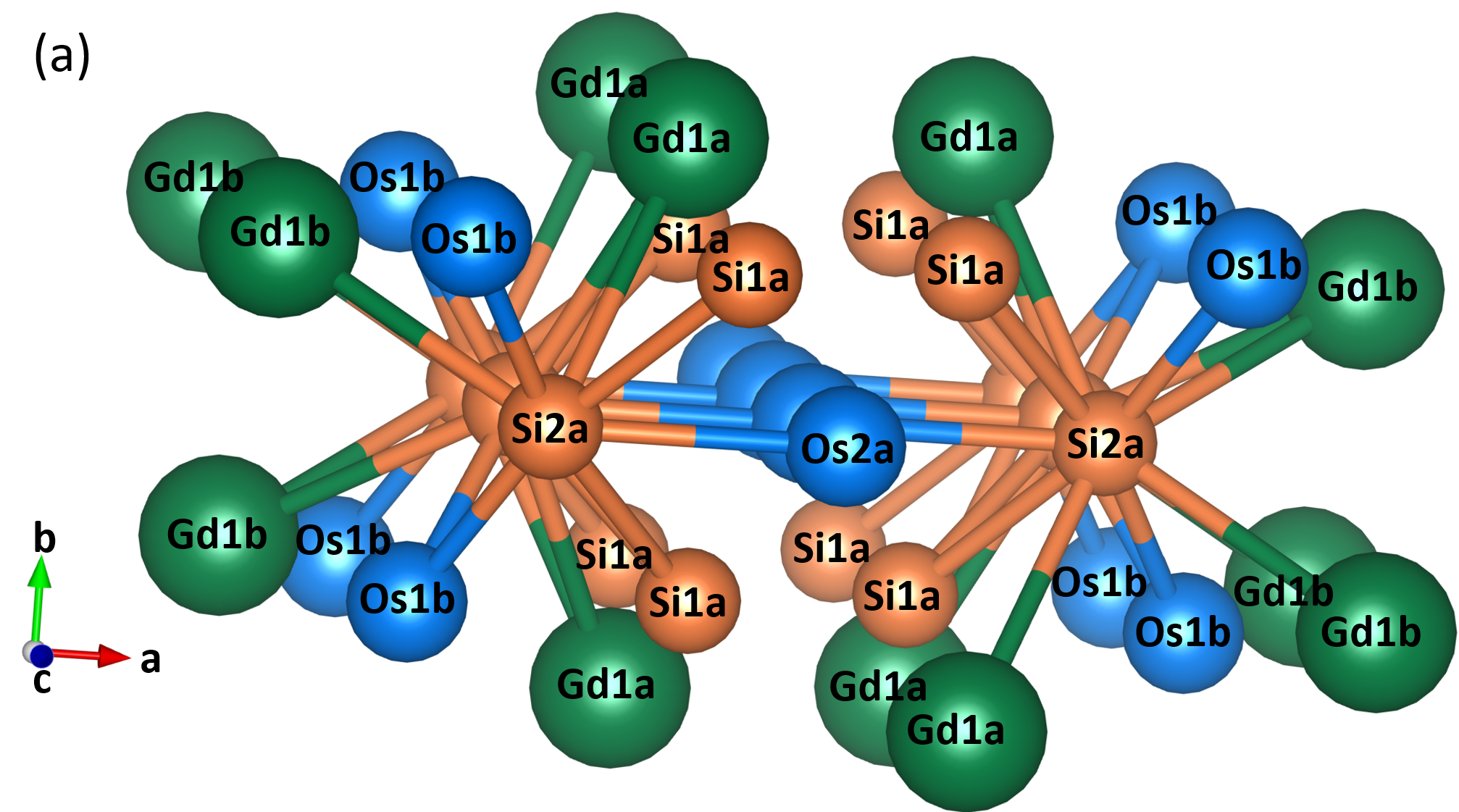}
\hspace{10mm}
\includegraphics[width=50mm]{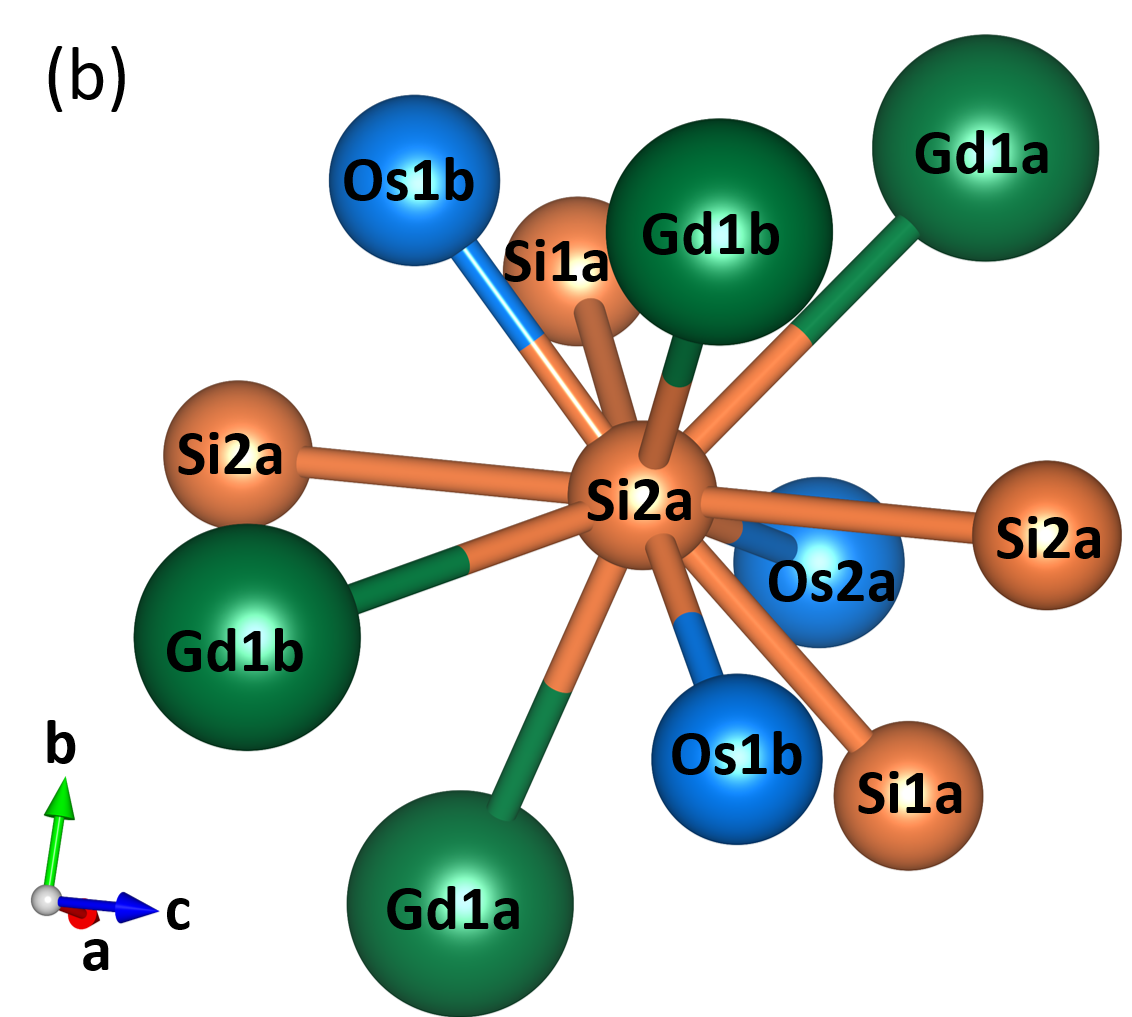}
\caption{\label{fig:gd2os3si5_cell2}%
(a) Extended view of the coordination of the Si2a atom.
(b) The eleven atoms in the first coordination sphere
of Si2a.
Neighboring atoms are
$2\times\,$Gd1a, $2\times\,$Gd1b, $2\times\,$Os1b,
$1\times\,$Os2a, $2\times\,$Si1a, $2\times\,$Si2a.
$t$-Plots of the variations of these distances in
the modulated structure are provided in
Fig. \protect\ref{fig:gd2os3si5_disttplot}.
}
\end{figure}
%***********************Figure8**********************************
%
while Si2b has small displacements.
This feature demonstrates the deviation from
tetragonal symmetry of the CDW modulation,
since tetragonal symmetry would imply equal
magnitudes of the amplitudes for Si2a and Si2b atoms.
Accordingly, the modulation mainly affects
bonding around Si2a atoms
(Fig. \ref{fig:gd2os3si5_cell2}).
Specifically, major variations are found for
the distances Gd1a--Si2a and Gd1b--Si2a,
and to a lesser extend for the distance Os1b--Si2a,
while the Os2a--Si2a bond is almost not modulated.
A large variation is found for the Si2a--Si2a
bonding contact (Fig. \ref{fig:gd2os3si5_disttplot}).

Specifically, large variations are found
for the four Si2a--Gd contacts, while the largest
variation is obtained for the distances to the two
Si2a neighbours of Si2a (Fig. \ref{fig:gd2os3si5_disttplot}).
The variation of contact distances is in phase for
three contacts Si2a, Gd1a and Gd1b, while they are
out of phase with the other group of three Si2a,
Gd1a and Gd1b atoms (Fig. \ref{fig:gd2os3si5_disttplot}).
%
%***********************Figure9*********************************
\begin{figure}[ht]
	\includegraphics[width=80mm]{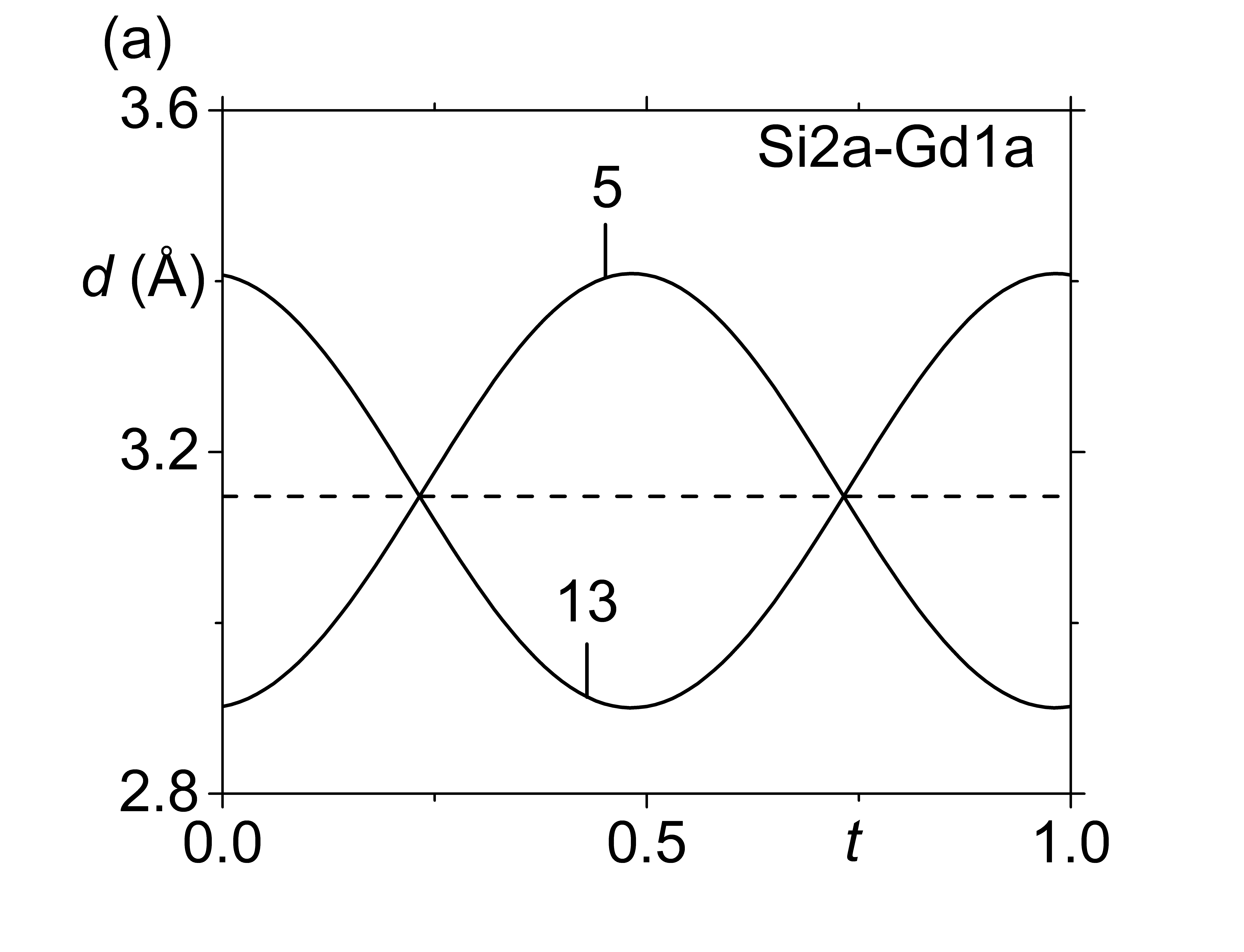}
	\includegraphics[width=80mm]{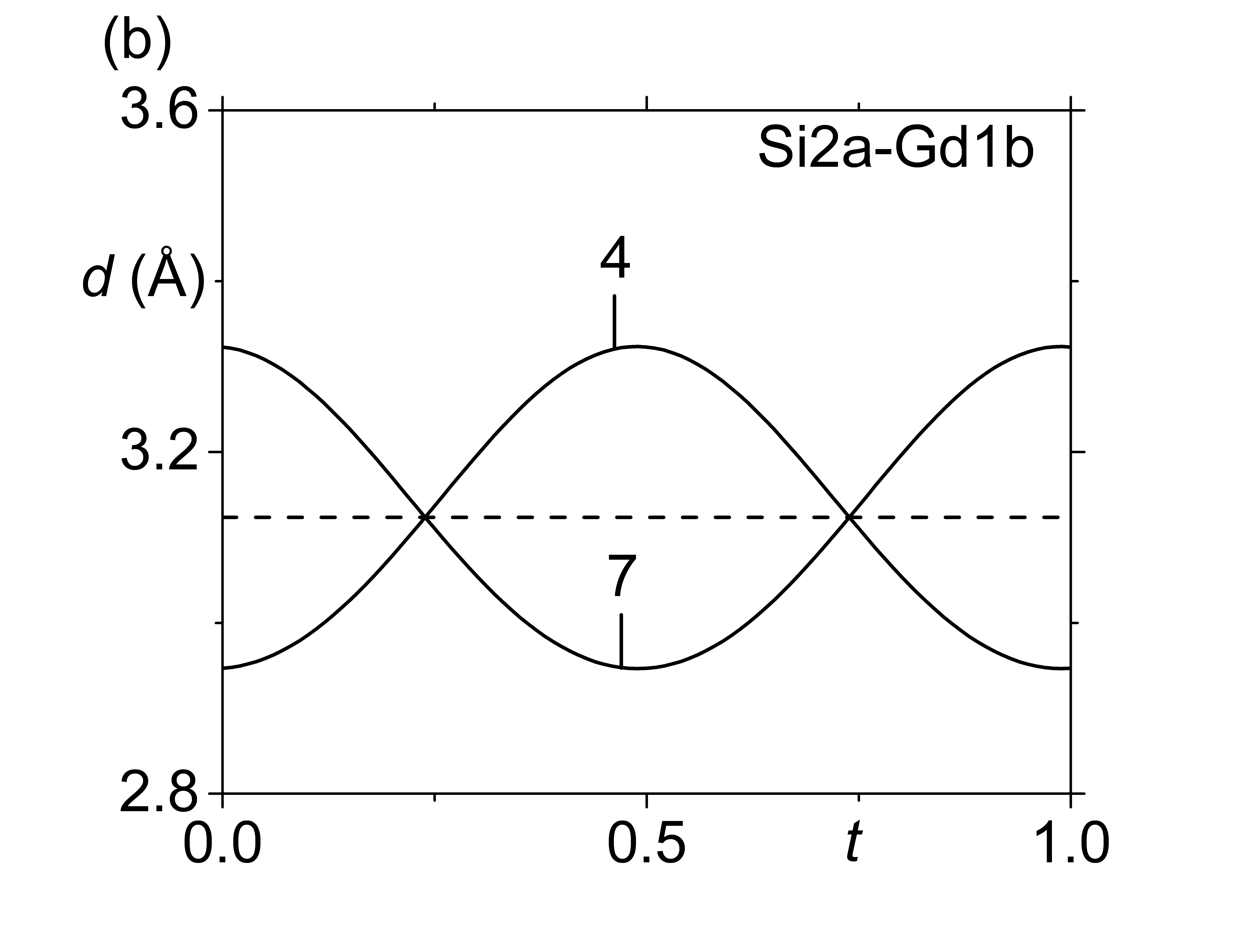}
	\hfill
	\includegraphics[width=80mm]{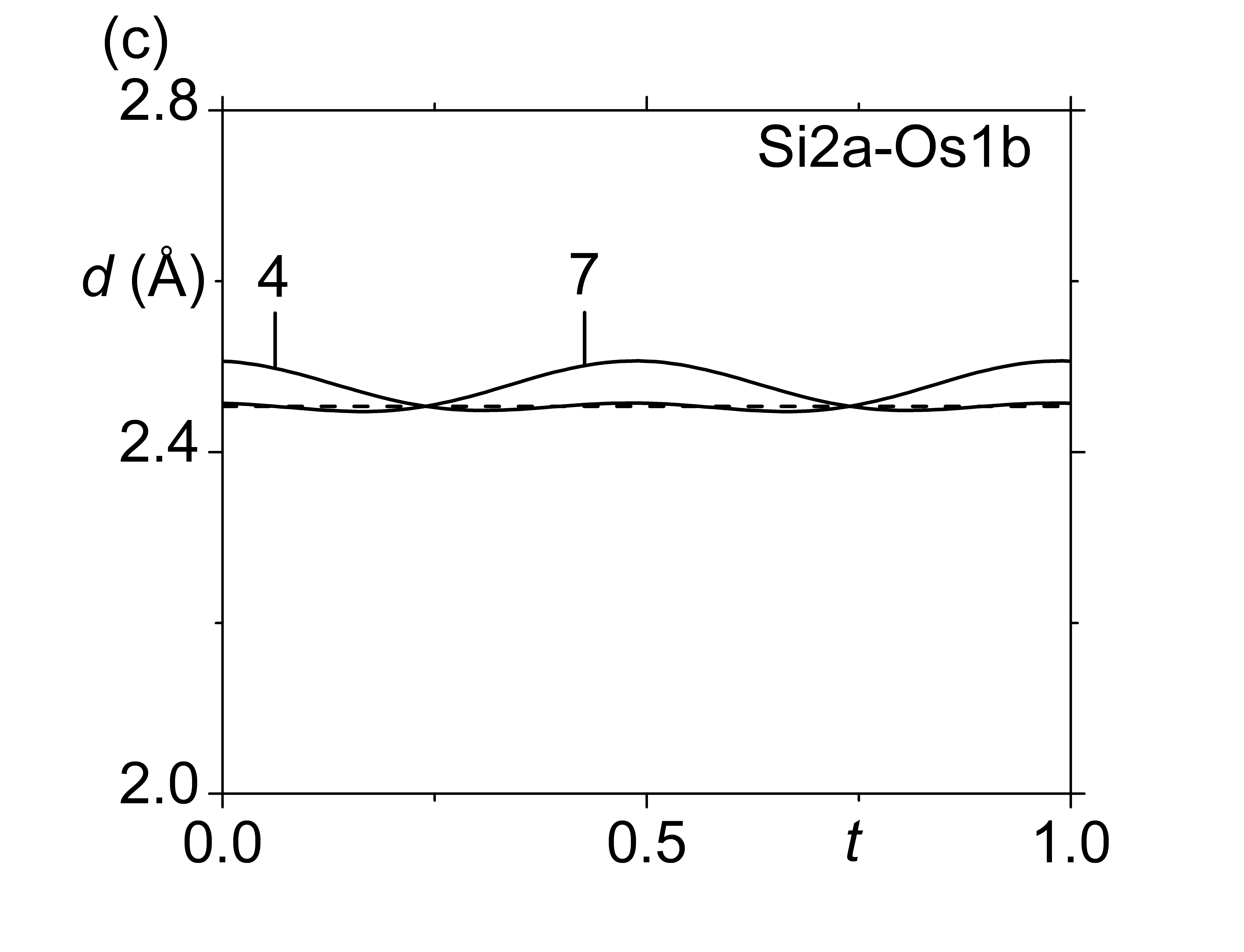}
	\includegraphics[width=80mm]{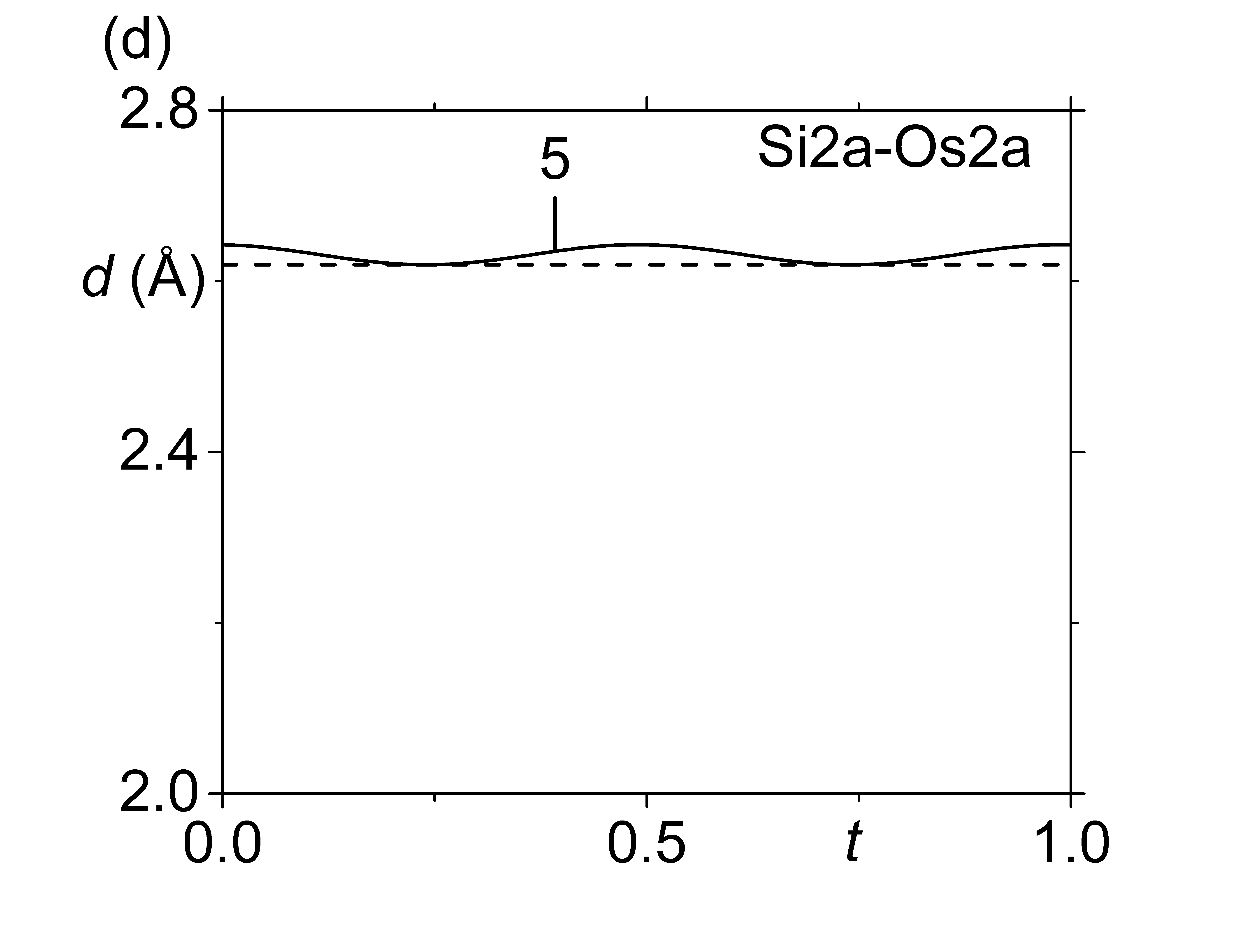}
    \hfill
	\includegraphics[width=80mm]{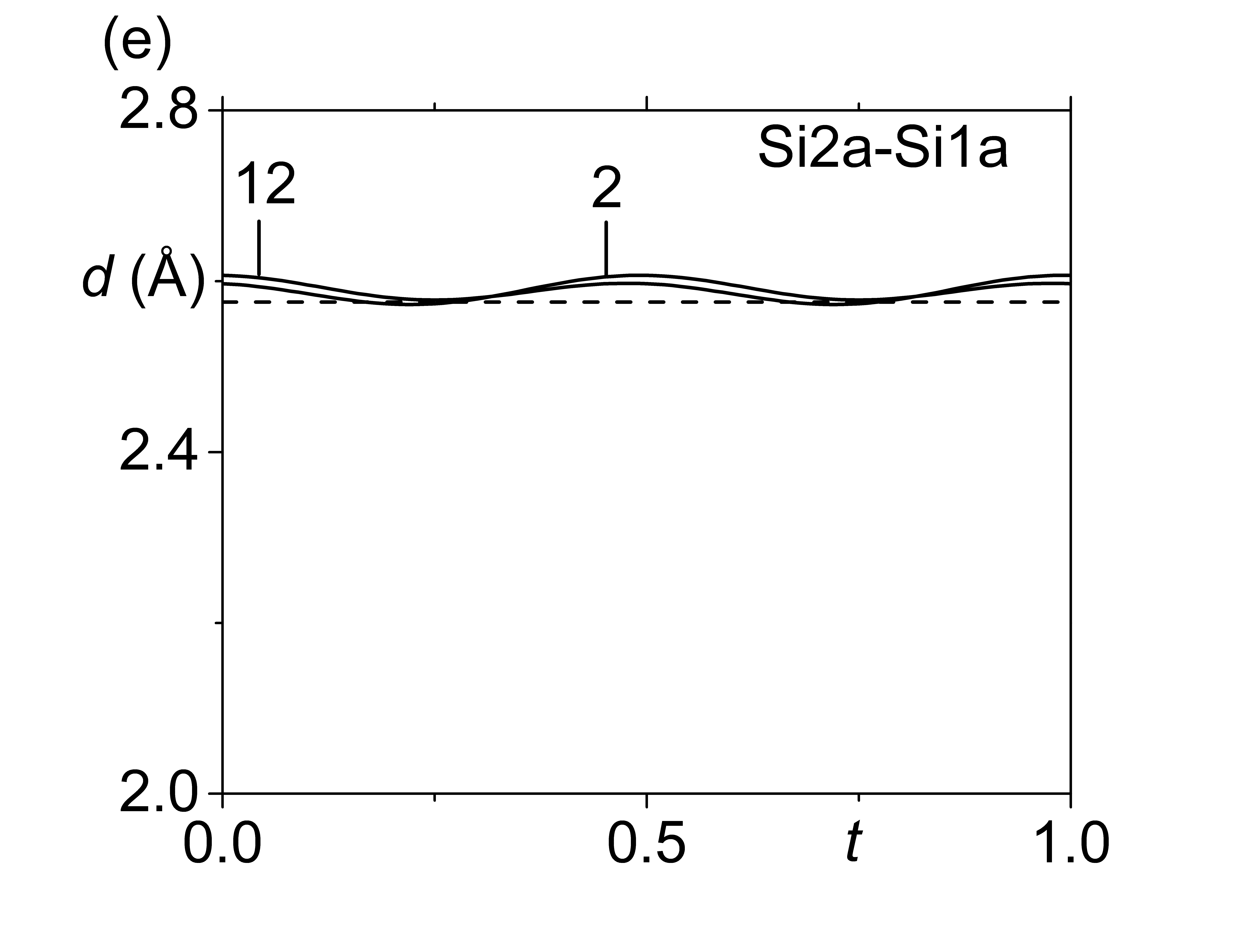}
    \includegraphics[width=80mm]{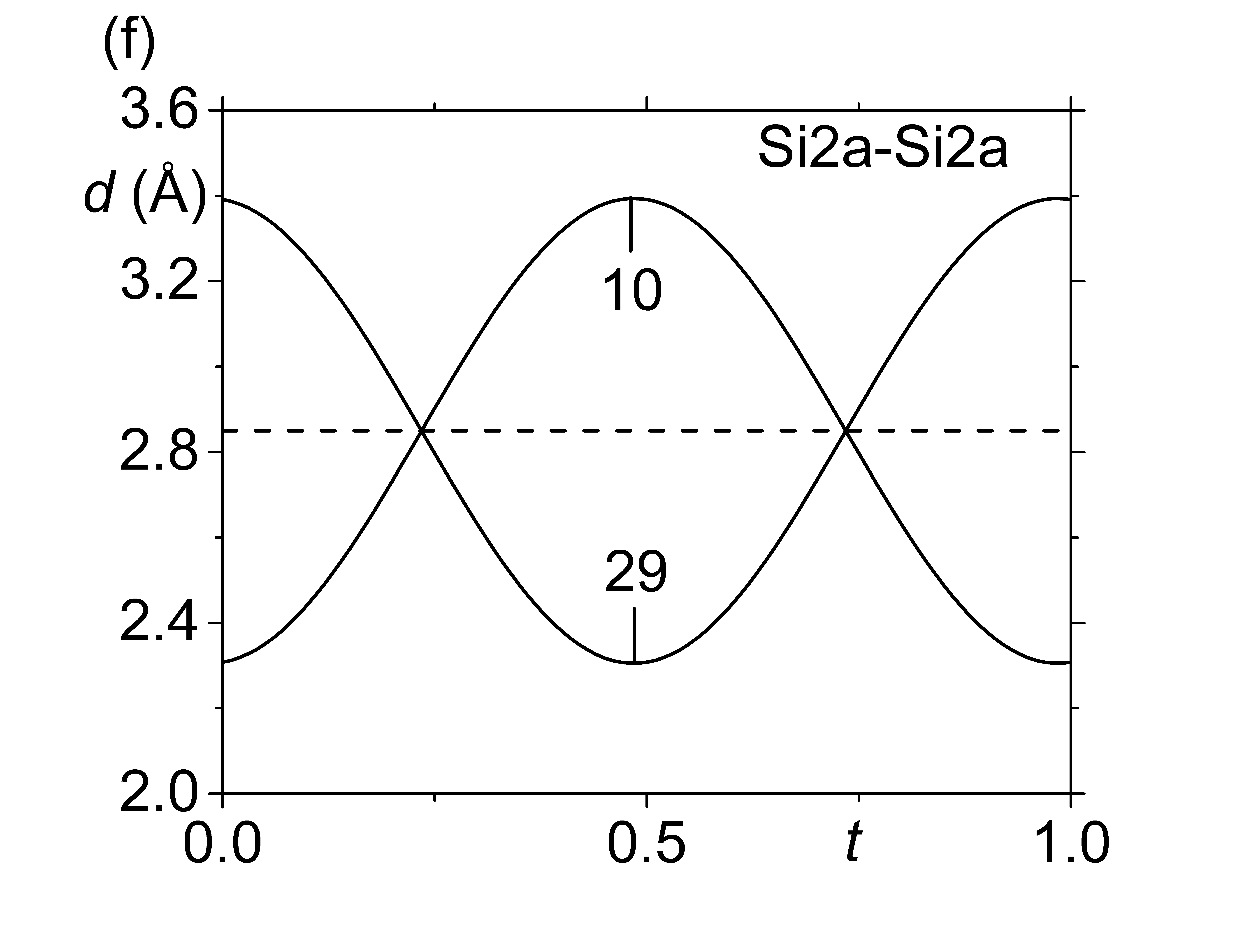}
\caption{\label{fig:gd2os3si5_disttplot}%
$t$-Plots of interatomic distances $d$ (\AA{}) of
Si2a--Gd1a, Si2a--Gd1b, Si2a--Os1b, Si2a--Os2a,
Si2a--Si1a and Si2a--Si2a at 200 K.
All plots refer to a single Si2a atom; compare to
Fig. \protect\ref{fig:gd2os3si5_cell2}.
$t$-Plots display interatomic distances as a function
of the phase $t$ of the modulation wave \cite{van2007incommensurate}.
The number on each curve is the number of
the symmetry operator that is applied to
the coordinating atom.
Symmetry operators are listed in Table S9 in
the Supplemental Material \cite{gd2os3si5suppmat2023a}.}
\end{figure}
%*********************Figure9***********************************
%
A complete set of $t$-plots of short interatomic
distances is given in the Supporting Information
\cite{gd2os3si5suppmat2023a}.
The gadolinium and osmium atoms are too far apart
from each other for a significant direct interaction
between them.
This indicates that the CDW deviates
from typical CDW systems, as they do not involve
metal-metal contacts, since the CDW in
Gd$_2$Os$_3$Si$_5$ is caused by Si2a atoms. This
is in contrast to the  $RE_{2}Ir_{3}Si_{5}$ systems
(R = Lu, Er, Ho) which are orthorhombic as the CDW
involves the Ir atoms and the Si atoms. For detailed
comparison of the CDW in the tetragonal and orthorhombic
variants refer to section S8 in the Supporting information
\cite{gd2os3si5suppmat2023a}.

\subsection{\label{sec:gd2os3si5_origin_cdw}%
Origin of the CDW in Gd$_2$Os$_3$Si$_5$}

As described above, the CDW in Gd$_2$Os$_3$Si$_5$
involves the modulation of Si atoms as opposed
to Ir--Ir zig-zag chains in Ho$_2$Ir$_3$Si$_5$ \cite{ramakrishnan2023a}
or Ge--Ge zig-zag chains in Sm$_2$Ru$_3$Ge$_5$ \cite{bugaris2017charge}.
A CDW can be stabilized by one of several mechanisms.\cite{pougetjp2024a}
These include Fermi surface nesting and hidden nesting \cite{gruner1994density,dressel_gruner_2002,johannes2008fermi,zhu2015classification,zhu2017misconceptions,whangbo1991hidden,roy2021quasi,pathak2022orbital}, wave vector-dependent electron-phonon coupling
\cite{weber2011extended,weber2013optical,maschek2015wave,maschek2018competing},
strong electron correlations \cite{chen2016charge,zhu2015classification,gerber2015three},
and a large electronic density of states (EDOS) near
the Fermi level ($E_{F}$) having degenerate branches
\cite{boring2000condensed,soderlind2019density}.

We have calculated the electronic band structure
of Gd$_2$Os$_3$Si$_5$, and thus obtained the Fermi surface,
and the Lindhard susceptibility in the static limit ($\omega = 0$),
with real part Re$\{\chi_0({\mathbf{q}},\omega)\}$ and
imaginary part Im$\{\chi_0({\mathbf{q}},\omega)\}$.
Figure~S5
shows the orbital- and atom-resolved electronic band
structure and EDOS.
As one can observe, the states near $E_{\rm F}$ are primarily
formed by Gd-$d$ and Os-$d$ orbitals with little
contribution from Si-$p$ orbitals.
Importantly, the EDOS at $E_{\rm F}$ is relatively low,
thus ruling out a large EDOS at $E_{\rm F}$ to be
the governing mechanism for CDW formation,
the latter which is quite
prevalent in the actinides \cite{boring2000condensed, soderlind2019density}.
Furthermore, we control the electron correlations
by increasing the Coulomb interaction parameter $U$
from 0 to 2~eV for Os atoms in calculations
which have significant contribution near $E_{\rm F}$,
but the band structure rigidly shifts by
$\sim$0.1\,eV with little influence on Fermi
surface topology and electron susceptibility
(Fig.S3).
Hence strong electron correlations are unlikely
to play governing role in the formation of
the CDW in Gd$_2$Os$_3$Si$_5$.

Next, we evaluate Fermi surface nesting and hidden nesting.
Figure \ref{fig:gd2os3si5_fermisurfaces}(a-d)
shows the Fermi surface where multiple electron
and hole pockets can be seen.
On visual inspection, some parallel contours
separated by a finite wave vector are visible
in Fig. \ref{fig:gd2os3si5_fermisurfaces}(b) and
 \ref{fig:gd2os3si5_fermisurfaces}(c).
We have calculated Im$\{\chi_0({\bf q},\omega)\}$
to identify whether the parallel contours will
lead to divergence, a signature of Fermi surface nesting mechanism \cite{gruner1994density,dressel_gruner_2002,johannes2008fermi,zhu2015classification,zhu2017misconceptions,whangbo1991hidden,roy2021quasi,pathak2022orbital}.
Figure \ref{fig:gd2os3si5_fermisurfaces}(e) displays  Im$\{\chi_0({\bf q},\omega)\}$ in the ($H$,$K$,0) scattering plane that shows the divergence near zero wave vectors only (i.e., at the corners); hence pointing towards the absence of finite wave vector CDW driven by Fermi surface nesting.
Moreover, since electronic bands disperse linearly near $E_{F}$
(Fig.~S4(b)),
it is possible, similar to $\alpha$-U and EuTe$_4$~\cite{roy2021quasi,pathak2022orbital},
that hidden nesting may drive the CDW transition here.
To explicitly confirm the role of hidden nesting, we further calculate Re$\{\chi_0({\bf q},\omega)\}$ as shown in Fig.~\ref{fig:gd2os3si5_fermisurfaces}(f).
We obtain weak divergence near zero wave vectors but no divergence at experimentally obtained CDW wave vector in Gd$_2$Os$_3$Si$_5$.
Note that, as discussed earlier, since the CDW involves modulation of Si atoms which have little contribution near $E_{\rm F}$, it is reassuring that our calculations explicitly confirm that nesting nor strong correlations of Gd and Os bands would control the formation of CDW.
From the above analysis, it seems likely that wave vector-dependent electron-phonon coupling is the dominant mechanism in Gd$_2$Os$_3$Si$_5$, where electronic states near $E_{\rm F}$ interact with a phonon branch involving Si displacements at CDW wave vector to induce phonon softening; and subsequently, the freezing of Si vibrations below $T_{\rm CDW}$ at CDW wave vector leads to the formation of CDW observed in experiments. Due to the large unit cell (40 atoms) and 100-orbital basis required to reproduce bands near $E_{\rm F}$, it remains challenging to confirm the wave vector-dependent electron-phonon coupling mechanism theoretically.

\subsection{\label{sec:gd2os3si5_discus_physical_properties}%
Physical Properties and the CDW phase transition}

\begin{figure*}
\includegraphics[width=0.85\textwidth]{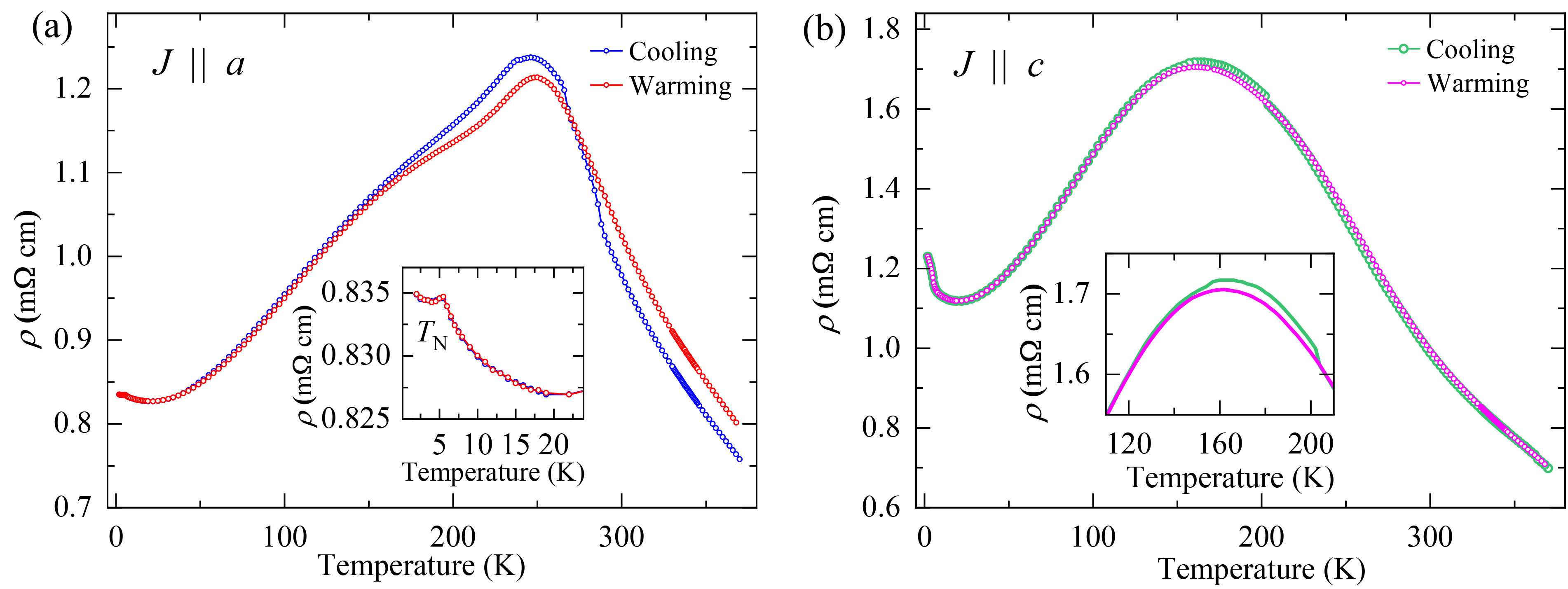}
\caption{\label{fig:gd2os3si5_electrical resistivity}%
Temperature dependent electrical resistivity during
cooling and warming cycles for
(a) $J_{\mathbf{a}}$ $\parallel$ $\mathbf{a}$, and
(b) $J_{\mathbf{c}}$ $\parallel$ $\mathbf{c}$,
insets show the expanded view at low temperatures.} 	
\end{figure*}

Figure~\ref{fig:gd2os3si5_electrical resistivity} shows the temperature dependence of electrical resistivity $\rho(T)$ for current ({\bf${J}$}) along the two principal crystallographic directions,  ${\bf a}$ and  ${\bf c}$,  for the cooling and
warming cycles in the temperature range  370 to 2~K.
%
%******************Figure 10*****************************
%*************************End of Fig.10*******************************
%
For $J_{\mathbf{a}}$, $\rho(T)$ increases with
decrease in the temperature down to 250~K for
both cooling and warming cycles, showing
semiconducting behavior.
$\rho(T)$ decreases with further decrease of
the temperature with change in the slope
near 220~K and a broad hump around 150~K.
There is a subtle increase in $\rho(T)$
below 20~K due to enhanced spin-disorder
scattering,
which is confirmed from $\chi(T)$ as discussed later.
$\rho(T)$ suddenly drops in the magnetically
ordered state below $T_{{\rm N}}~\sim$~5.5~K.
This shows a metallic behavior in the temperature range 250 to 2~K,
and semiconducting behavior between 370 and 250~K.
Note that the cooling data exhibits higher resistivity
 than the warming cycle between 100 and 270~K.
Above 270~K, the warming data obtains higher resistivity
values.
As a result, there are two hysteresis loops,
for 100 to 270~K and for 270  to 370~K and above,
 in the cooling and warming runs
[Fig. \ref{fig:gd2os3si5_electrical resistivity}(a)].
Below 100~K, both cooling and warming curves
more or less collapse onto a single curve along
with a broad hump around 150~K.

For $J_{\mathbf{c}}$, $\rho(T)$ increases
with decreasing temperature down to $\sim$ 170~K,
showing semiconducting behavior.
Further decrease in the temperature between
150 to 20~K, $\rho(T)$  decreases.
It shows a broad hump at around 160~K.
Unlike the case of $J_{\mathbf{a}}$,
$\rho(T)$ below $T_{{\rm N}} \sim$ 5.5~K
increases with decrease in the temperature,
which may be  attributed to the magnetic superzone gap.
We can see a significant hysteresis between
cooling and warming runs between 140 to 205~K
(inset of Fig~\ref{fig:gd2os3si5_electrical resistivity}(b)).
These warming and cooling curves do not collapse
onto a single curve for temperatures
up to 370~K, while a hysteresis is observed.
From the $\rho(T)$ analysis, we can state that
there are two main features:
(i) insulator to metal transition (IMT), and
(ii) hysteresis in cooling and warming cycles
in the metallic as well as the insulating regimes.
The IMT along with hysteresis appears
to indicate CDW phase that is in line
with observed superlattice peaks in SXRD.
Hysteresis in cooling and warming cycles
in the metallic regime may associate with
the structural phase transition.
However, we did not find any change in
the modulated structure below 300~K  from the SXRD data.
At the same time, superlattice reflections are
pronounced in the hysteretic regime, which
suggests that there may be successive CDW
states in this compound.

\begin{figure*}
\includegraphics[width=0.8\textwidth]{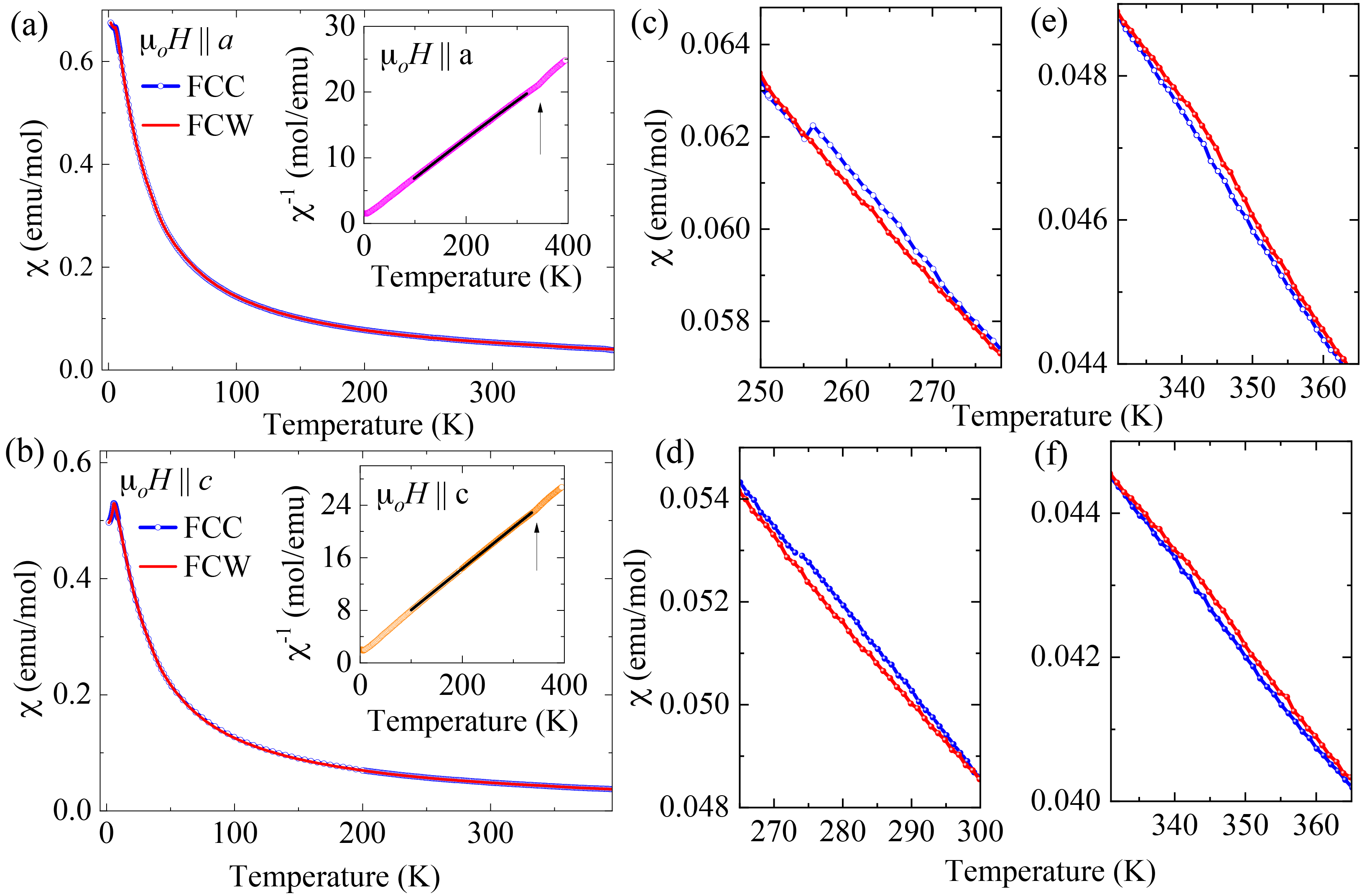}
\caption{\label{Fig:gd2os3si5_fig3}%
(a) Temperature dependent magnetic susceptibility
during FCC and ZFC runs of Gd$_2$Os$_3$Si$_5$ for
(a) $\mu_0H$ $\parallel$ $\mathbf{a}$ and
(b) $\mu_0H$ $\parallel$ $\mathbf{c}$ at 50 Oe.
Expanded view of susceptibility at different
temperature regimes for
(c, e) $\mu_0H$ $\parallel$ $\mathbf{a}$ and
(d, f) $\mu_0H$ $\parallel$ $\mathbf{c}$.
Insets (a, b) show Curie-Wiess (CW) fitting
in the temperature range 100-320~K.}	
\end{figure*}

The temperature-dependent magnetic susceptibility
$\chi(T)$ has been measured during field-cooled
cooling (FCC) and
zero-field-cooled (ZFC) protocols,
in an applied magnetic field of $H = 50$ Oe
for the field parallel to $\mathbf{a}$
($\chi_a(T)$) and
parallel to $\mathbf{c}$ ($\chi_c(T)$)
[Fig. \ref{Fig:gd2os3si5_fig3}(a,b)].
In FCC, the data were recorded during cooling cycle of the
sample from 390 to 1.8~K.
In ZFC, the data were recorded during
warming cycle from 1.8 to 390~K,
after cooling the sample in zero field.
The magnetic susceptibility exhibits anomalies in
both $\chi_a(T)$ and $\chi_c(T)$ at $T_{{\rm N}} \simeq$ 5.5~K,
due to long-range antiferromagnetic order below $T_{\rm N}$.
Above $T_{{\rm N}}$, $\chi_a(T)$ and $\chi_c(T)$ decreases
with increasing temperature, with no sign of anisotropy
as expected for a $S$-state Gd-ion.

We have obtained a good fit of the Curie-Weiss (CW)
law to both the data $\chi_a(T)$ and the data
$\chi_c(T)$ of the ZFC protocol,
in the temperature range $100-320$~K
(Fig.~\ref{Fig:gd2os3si5_fig3}(a,b), insets).
%
%******************Figure11**************************
%******************END of Figure 11*********************
%
These fits resulted in $\mu_{\rm eff}$ = 7.9~$\mu_{\rm B}$/Gd
for both the directions, a value which is close to the
theoretical value of 7.94~$\mu_{\rm B}$/Gd for Gd$^{3+}$.
Interestingly, there is a change in slope of the
$\chi^{-1}(T)$ \textit{vs} $T$ plots at $T = 345$~K,
which we interpret as the signature of a CDW phase transition.
%The CW law deviates above 330~K due an anomaly
%near 345~K in $\chi$ for both the directions
%(Fig.~\ref{Fig:gd2os3si5_fig3}(a,b), insets).

The magnetic susceptibility exhibits hysteresis for
FCC and ZFC runs between 255 and 270~K for $\chi_a(T)$,
and between 265 and 300~K for  $\chi_c(T)$
[Fig.~\ref{Fig:gd2os3si5_fig3}(c,d)].
Such a hysteretic behavior is also seen in $\rho(T)$ in
cooling and warming cycles for $J_a$
(Fig.~\ref{fig:gd2os3si5_electrical resistivity}(a).
A second measurement on the same sample led to a
hysteresis in $\chi_a(T)$ of 260--285~K
[the inset in Fig. S2(a)],
and of 335--370~K (the inset in Fig. S2(b)].
The slightly different temperature ranges are
attributed to different thermal histories
of the sample, when the two measurements were performed.

Notably, the inverse magnetic susceptibility exhibits
a change in the slope near 345~K for both $\chi_a(T)$
and $\chi_c(T)$ [inset in Fig.~\ref{Fig:gd2os3si5_fig3}(a,b)].
An expanded view of $\chi_a(T)$ and $\chi_c(T)$
reveals a small hysteresis for 330 to 365~K
[Fig.~\ref{Fig:gd2os3si5_fig3}(e,f)].
We believe that this high temperature anomaly at
$T_{\rm CDW}$ $\sim$ 345~K is associated with the CDW
phase transition.
This value for $T_{\rm CDW}$ is in agreement with the
results from high-temperature SXRD
(Section \ref{sec:gd2os3si5_incommensurate_structure}).

%*********************************End****Table4************
%

\subsection{\label{sec:gd2os3si5_criterion_transitions}%
Criterion for the formation of CDW/structural
transitions in $RE_{2}T_{3}X_{5}$ compounds}

Compounds $RE_{2}T_{3}X_{5}$ ($RE$ = rare-earth,
$T$ = transition metal, $X$ = Si, Ge) with either
the U$_2$Co$_3$Si$_5$ or the Sc$_2$Fe$_3$Si$_5$
structure type may or may not develop a CDW.
We have found that CDW formation in these compounds
can be related to their lattice parameters,
according to the value of $c/\sqrt{ab}$.

We have compiled a list of $2\:3\:5$ materials
from the handbook on 2:3:5 materials \cite{brown2023a}
and other sources \cite{kyrktm2022a, godart1988a, anand2007b,
naka2015b, frecc2021a, anand2008a, kurenbaeva1998a,
frecc2019a, becker1997a, hoss2000a, gam2017a,
li2019a, bug2017b, mazumdar1992a, kama2020a,
mazumdar1997a, mazumdar2003a, mazumdar1999a,
hashi2007a, mazumdar1994a, huo2002a, muro2008a,
anand2012a, kuren1999a, nirmala2000a, griv2013a,
sarkar2015a, bauer2011a, skan1997a, mazumdar1996a,
singh2004a, patil1997a, ramakrishnan2001a,
paccard1987a, chevalier1982a, paccard1990a,
gribanov2010a, rizzoli2004a, singh2002a, watanabe2009a,
vining1983a, braun1981a, segre1981a, jasper1996a,
wastin1994a, morozkin1998a, morozkin1998b, nirmala2001a,
nirmala2001b, meyer1985a, chabot1984a, chabot1985a},
along with their values of $c/\sqrt{ab}$.
Compounds with a monoclinic structure are not included.
Also, we do not consider 2:3:5 stannides or gallides,
as their lattice parameters are completely different
from compounds with the
U$_2$Co$_3$Si$_5$ or Sc$_2$Fe$_3$Si$_5$ structure types \cite{brown2023a}.
Currently, we expect that $c/\sqrt{ab}$ falls
between 0.5260 to 0.5432 for those compounds
that undergo CDW and/or structural phase transitions.
Basically we consider five scenarios.
\begin{description}
\item
Case 1: Confirmed CDW/structural transition and fulfilled criterion.
\item
Case 2: Confirmed CDW/structural transition and outside criterion.
\item
Case 3: Confirmed absence of CDW/structural transition (e.g. by temperature-dependent
physical properties which are published) and fulfilled criterion
(i.e. outside the CDW/structural transition range).
\item
Case 4: Confirmed absence of CDW/structural transition and violated criterion.
\item
Case 5: No CDW/structural transition reported, but experiments have not been done
that could confirm the presence/absence of a CDW.
\end{description}

\begin{figure}[ht]
\includegraphics[width=0.75\textwidth]{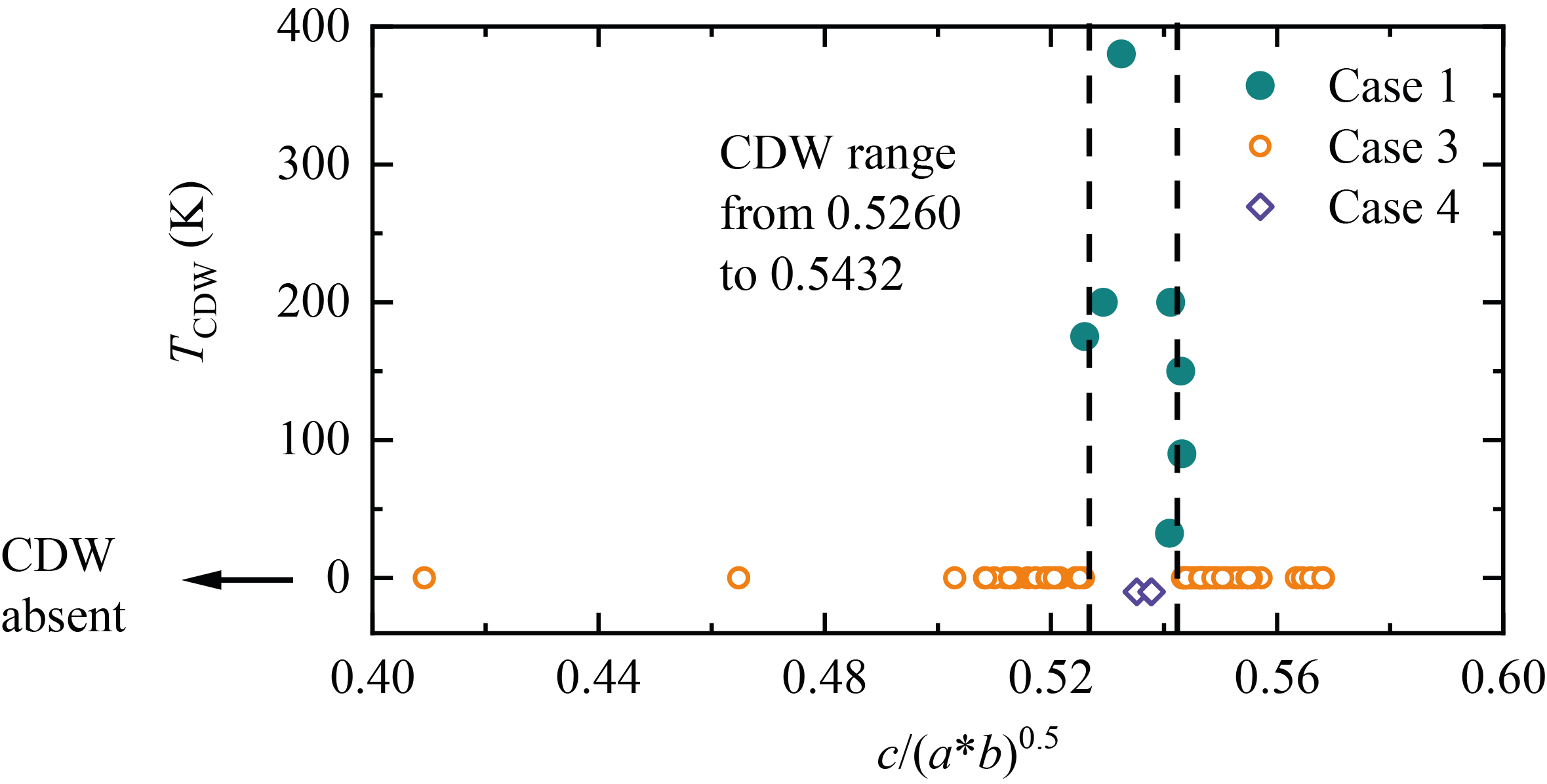}
\caption{\label{tab:gd2os3si5_case1_4}%
Plot showing $T_{\rm CDW}$ vs $c/\sqrt{ab}$ for cases 1,
3 and 4.
There are no materials which qualify for case 2.}
\end{figure}

A CDW or any other structural transition is likely
to proceed at some temperature,
if the $c/\sqrt{ab}$ lies in the range: $0.5260 < c/\sqrt{ab} < 0.5432$, while compounds
with $c/\sqrt{ab}$ outside this range do not have
a phase transition (Fig. \ref{tab:gd2os3si5_case1_4}).
%
%*********************************Fig12**********************
%***************************End of Fig.12*******************
%
There are three materials that fulfil the criterion,
but do not form a CDW (case 4).
They are La$_2$Rh$_3$Si$_5$, Lu$_2$Rh$_3$Si$_5$
and Yb$_2$Pd$_3$Ge$_5$ \cite{ramakrishnan2001a,frecc2019a}.
The absence of a phase transition  might be related to
a large number of defects, as they can be found in
polycrystalline materials.
Suppression of CDW is known to occur in materials
with chemical disorder or defects like vacancies.
Examples include CuV$_2$S$_4$, NiV$_2$Se$_4$,
ErTe$_3$ \cite{ramakrishnan2019a,ramakrishnan2023b,fang2019a}.

%*********************************Fig13**********************
\begin{figure}[ht]
\includegraphics[width=0.75\textwidth]{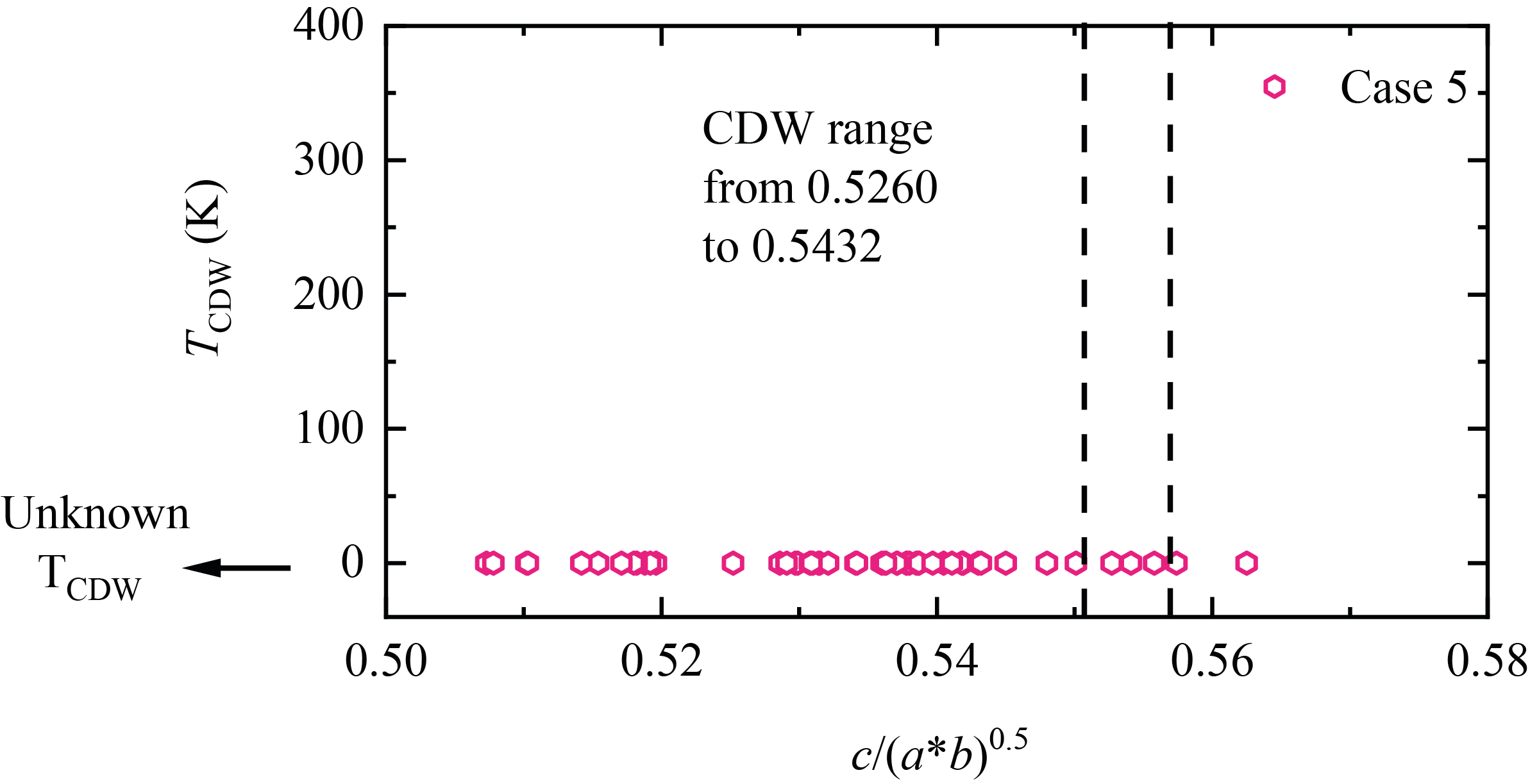}
\caption{\label{tab:gd2os3si5_case5}%
Plot showing $T_{\rm CDW}$ vs $c/\sqrt{ab}$ for case 5.}
\end{figure}
%**********************End of Fig13********************

Figure \ref{tab:gd2os3si5_case5} collects all materials where there
are no structural transitions reported due to lack of evidence
from measurements.
Many of these compounds that fall within
the criterion are the $RE_2$Rh$_3$Si$_5$ series which are
reported to exhibit AFM transitions from the
physical properties \cite{ramakrishnan2001a}.
However, no diffraction experiments have been done to see the AFM
is coupled to the lattice to induce a structural distortion.
The observed criterion can be used for the prediction
of compounds that might form a CDW
(Fig. \ref{tab:gd2os3si5_case5}).
These compounds are listed as case 5 in
table S8 in the supporting information. This table provides a complete list of materials used in the discussion \cite{gd2os3si5suppmat2023a}.

%\clearpage

\section{\label{sec:gd2os3si5_conclusions}%
Conclusions}

We have synthesized a new polymorph of
Gd$_2$Os$_3$Si$_5$ and established
that it is a high temperature CDW compound.
The CDW transition occurs below 400~K according
to the SXRD measurements, presumably close to 345~K
as indicated by the anomalies detected in the magnetic
susceptibility.
The CDW crystal structure of the modulated phase
indicates that there is a lowering from
tetragonal ($P$4/$mnc$) to orthorhombic symmetry ($Cccm(\sigma00)0s0$)
at the transition.
There is no distortion in the lattice away from
tetragonal symmetry as the modulation
solely causes the reduction in the overall symmetry.
The CDW appears to be primarily
influenced by the Si2a atoms due to the large
modulation displacements of these atoms, also
in agreement with Sm$_2$Ru$_3$Ge$_5$, as the
Ge atoms have been reported to possess the largest
modulation amplitudes \cite{bugaris2017charge}.
Electronic structure simulations rule out the Fermi surface
nesting and hidden nesting, strong electron correlations,
and a large electronic density of states (EDOS) near
the Fermi level ($E_{F}$) having degenerate branches as the
possible origin of CDW formations and suggest wave
vector-dependent electron-phonon coupling to be
the governing mechanism.
Lastly, we also reported that CDW
in the 2:3:5 systems are influenced by a criterion
which is based on the value of $c/\sqrt{ab}$
being between 0.5260 and 0.5432.
This criterion can be useful to predict possible
new CDW compounds in the 2:3:5 family.

\section{Supporting information}

The Supporting Information is available free of charge at [URL to be inserted by the publisher].

1. Details of the SXRD data collection, data processing, structural analysis and DFT (PDF). \\
2. CIF of the periodic and periodic and incommensurately modulated structures at various temperatures.

\begin{acknowledgement}
We thank C. Paulmann for the assistance in
collecting SXRD data at Beamline P24.
We thank Dr. C. D. Malliakas and
Prof. Dr. M. Kanatzidis for providing the
SXRD data of the modulated structure
of Sm$_2$Ru$_3$Ge$_5$.
We acknowledge DESY (Hamburg, Germany), a member of
the Helmholtz Association HGF, for the provision
of experimental facilities.
Parts of this research were carried out at
PETRA III, using beamline P24.
Beamtime was allocated for proposal I-20220188.
\end{acknowledgement}

%\bibliography{gd2os3si5}
\providecommand{\latin}[1]{#1}
\makeatletter
\providecommand{\doi}
  {\begingroup\let\do\@makeother\dospecials
  \catcode`\{=1 \catcode`\}=2 \doi@aux}
\providecommand{\doi@aux}[1]{\endgroup\texttt{#1}}
\makeatother
\providecommand*\mcitethebibliography{\thebibliography}
\csname @ifundefined\endcsname{endmcitethebibliography}
  {\let\endmcitethebibliography\endthebibliography}{}

\clearpage

\end{document}